\documentclass[12pt,aps,prb,preprint]{revtex4}
\usepackage{amsmath,amssymb,amsfonts,mathrsfs} 
\usepackage{graphicx}   
\usepackage{bm}

\def\bp{\mathbf p}

\def\bP{\mathbf P}
\def\bB{\mathbf B}

\def\br{\mathbf r}

\def\dbp{\Delta\mathbf p}
\def\bm{\mathbf m}
\def\br{\mathbf r}

\begin{document}

\title{On electromagnetic momentum of an electric dipole in a magnetic field}

\author{Ben Yu-Kuang Hu}
\email{byhu@uakron.edu} \affiliation{Department of Physics,
University of Akron, Akron, OH~44325-4001}

\date{\today}

\begin{abstract}

The total linear electromagnetic field momentum $\mathbf P_{\mathrm{em}}$ of a stationary electric dipole $\mathbf p$ in a static magnetic field $\mathbf B$ is considered.
The expression $\mathbf P_{\mathrm{em}} = \frac12\mathbf B \times \mathbf p$, which has previously been implied to hold in all static magnetic field situations, is not valid in general. The contribution of the electromagnetic momentum of the fringing fields of the dipole is discussed.    It is shown that when either the static magnetic field or the electric dipole moment is changed, the mechanical impulse on the system equals $-\Delta\mathbf P_{\mathrm{em}}$, and hidden momentum does not need to be invoked in order to conserve total momentum.
\end{abstract}
\maketitle

\section{Introduction} \label{sec:I}

In classical electromagnetism, the electric ($\mathbf E$) and magnetic ($\mathbf B$) fields store linear momentum in the form of electromagnetic momentum density\cite{griffiths,jackson} $\epsilon_0 (\mathbf E \times \mathbf B)$ [SI units are used throughout this paper].  The total electromagnetic momentum of a system is the integral of the electromagnetic momentum density,
\begin{equation}
\mathbf P_{\mathrm{em}} = \epsilon_0 \int \mathbf E(\mathbf r) \times \mathbf B(\mathbf r)\ d\mathbf r.\label{eq:Pem}
\end{equation}  It is therefore possible for static electric and magnetic fields to store electromagnetic momentum (although the {\em total} momentum of the system must zero, as is discussed later).  To illustrate the consequences of the electromagnetic momentum on a specific system, several authors have studied the electromagnetic momentum due to electric dipoles in the presence of static magnetic fields.\cite{mcdonald,gsponer,babson}

In this paper, we consider two aspects of electromagnetic momentum stored in the fields due to a stationary electric dipole in a static magnetic field.  The first is the expression for the electromagnetic momentum stored in the system, which has been reported to be (see Refs.~\onlinecite{mcdonald,gsponer,babson} and Ref.~\onlinecite{griffithsresource})
\begin{equation}
\mathbf P_{\mathrm{em}} = \frac12 \mathbf B  \times \mathbf p,\label{eq:P=Bxp}
\end{equation}
where $\mathbf p$ is the electric dipole and $\mathbf B$ is the magnetic field at the position of dipole.  Eq.~(\ref{eq:P=Bxp}) was derived\cite{mcdonald,gsponer,babson} for stationary electric dipoles in locally uniform static magnetic fields produced by a long uniform solenoid and a spinning sphere with a uniform surface charge density.  Ref.~\onlinecite{griffithsresource} inadvertently omitted to mention that the result was derived for locally uniform fields, implying that it was true in general. 
Given that the electromagnetic momentum for a stationary magnetic moment $\mathbf m$ in a static (not necessarily uniform) electric field is\cite{griffithsresource,furry}
\begin{equation}
\mathbf P_{\mathrm{em}} = \frac1{c^2}\mathbf E \times \mathbf m,\label{eq:P=Exm}
\end{equation}
where $\mathbf E$ is the electric field at the position of the magnetic moment and $c$ is the speed of light, it seems plausible that there should be an equivalent expression for a stationary electric dipole in a static, non-uniform magnetic field.   However, the results Eqs.~(\ref{eq:P=Bxp}) and (\ref{eq:P=Exm}) are incompatible with each other, as illustrated in the following example.

\begin{figure}
\includegraphics{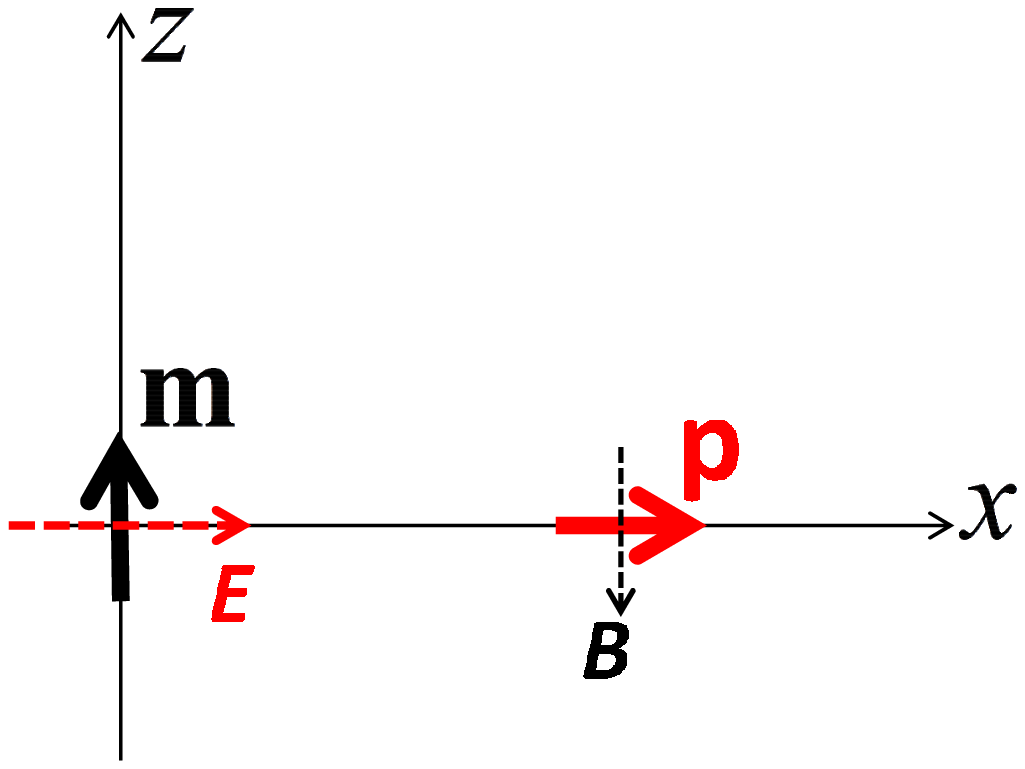}
\begin{caption}  {Configuration of electric ($\mathbf p$) at the origin and magnetic ($\mathbf m$) dipole at $R\hat{\mathbf x}$ which demonstrates that Eqs.~\ref{eq:P=Bxp} and \ref{eq:P=Exm} give inconsistent results for $\mathbf P_{\mathrm{em}}$ .   The dotted line labeled $\mathbf E$ is the electric field due to $\mathbf p$ at $\mathbf m$, and the dotted line labeled $\mathbf B$ is the magnetic field due to $\mathbf m$ at $\mathbf p$.    \label{fig:1}}
\end{caption}
\end{figure}

Assume there is a magnetic moment $\mathbf m = m \hat{\mathbf z}$ at the origin and an electric dipole $\mathbf p = p \hat{\mathbf x}$ at $R\hat{\mathbf x}$ (where carets indicate unit vectors), as shown in Fig.~\ref{fig:1}.  The magnetic field at a displacement $\mathbf r\ne 0$ from the magnetic moment is (see {\em e.g.}, Ref.~\onlinecite{griffiths}, p. 255)
\begin{equation}
\mathbf B = \frac{\mu_0}{4\pi r^3}[3(\mathbf m \cdot\hat{\mathbf r})\hat{\mathbf r} - \mathbf m].\label{eq:dipoleB}
\end{equation}
The electric field at a displacement $\mathbf r$ from the electric dipole is obtained by replacing $\mathbf m$ by $\mathbf p$ and $\mu_0$ by $\epsilon_0^{-1}$ in the above equation.
The magnetic field at $R\hat{\mathbf x}$ due to the magnetic dipole is $\mathbf B(\mathbf r = \hat{\mathbf x}R) = -\mu_0 m/(4\pi R^3)\,\hat{\mathbf z}$, and the electric field at the origin due to the electric dipole\cite{griffiths} is $\mathbf E(\mathbf 0) = p/(2\pi\epsilon_0 R^3)\,\hat{\mathbf x}$.   These together with Eqs.~(\ref{eq:P=Bxp}), (\ref{eq:P=Exm}) and $c^{-2} = \mu_0 \epsilon_0$ give
\begin{subequations}
\begin{align}
\mbox{From Eq.~(\ref{eq:P=Bxp}):}&\quad \mathbf P_{\mathrm{em}} = - \frac18\frac{\mu_0mp}{\pi R^3}\hat{\mathbf y};\\
\mbox{From Eq.~(\ref{eq:P=Exm}):}&\quad \mathbf P_{\mathrm{em}} = -\frac12\frac{\mu_0 mp}{\pi R^3}\hat{\mathbf y}.
\end{align}
\end{subequations}
Since $\mathbf P_{\mathrm{em}}$ evaluated using Eqs.~(\ref{eq:P=Bxp}) and (\ref{eq:P=Exm}) do not agree, at least one of these expressions is not valid in general.\cite{franklin}   It turns out that it is Eq.~(\ref{eq:P=Bxp}) that is not in general valid, even in cases where the magnetic field is locally uniform,   and $\mathbf P_{\mathrm{em}}$ cannot in general be written solely in terms of $\mathbf p$ and $\mathbf B$ at the position of the dipole.  In Section \ref{sec:II} of this paper, we give several expressions for $\mathbf P_{\mathrm{em}}$ for a stationary electric dipole in a static magnetic field, all of which depend (explicitly or implicitly) on the electric current configuration that generates the magnetic field. This is followed by examples and a discussion regarding the reasons behind the differences in total amount of $\mathbf P_{\mathrm{em}}$ for different current configurations.

Section \ref{sec:III} deals with another aspect of an electric dipole in a magnetic field: the impulse imparted to the system when either the electric dipole moment or magnetic field is reduced to zero.   A recent paper by Babson {\em et al.}\cite{babson} seemed to imply (although it was not their intent\cite{griffithsprivate}) that it is necessary to take into account the hidden momentum\cite{hiddenmomentum} in the system in order for the total momentum to be conserved.   We show that in these systems, the loss of electromagnetic momentum is always equal to the mechanical impulse imparted to the system, and it is not necessary to invoke the presence of hidden momentum to conserve the total momentum.

Section \ref{sec:IV} contains a discussion of the results of this paper.


\section{$\mathbf P_{\mathrm{em}}$ for a stationary electric dipole in a static magnetic field}\label{sec:II}

The total electromagnetic momentum $\mathbf P_{\mathrm{em}}$ for the case in which the electric and magnetic fields $\mathbf E$ and $\mathbf B$ are due to stationary charges distribution $\rho(\mathbf r)$ and steady current distributions $\mathbf J(\mathbf r)$ which are local (do not extend to infinity), can also be expressed as\cite{furry}
\begin{subequations}
\begin{align}
\mathbf P_{\mathrm{em}}& = \frac1{c^2} \int V(\mathbf r)\;\mathbf J(\mathbf r)\ d\mathbf r,\label{eq:PemVJ}
\end{align}
where $V(\mathbf r)$ is the scalar potential in the Coulomb gauge,
\begin{align}
V(\mathbf r) &= \frac1{4\pi\epsilon_0} \int \frac{\rho(\mathbf r')}{\vert\mathbf r - \mathbf r'\vert}\ d\mathbf r'.\label{eq:V}
\end{align}
\end{subequations}
In Appendix \ref{appendix:A}, the derivation of Eq.~(\ref{eq:PemVJ}) is reproduced, and circumstances in which the locality of $\mathbf J$ can be relaxed are discussed.

The electromagnetic momentum can alternatively be expressed as\cite{furry,thomson,calkin}
\begin{subequations}
\begin{align}
\mathbf P_{\mathrm{em}}& = \int \rho(\mathbf r)\ \mathbf A(\mathbf r)\ d\mathbf r.\label{eq:PemrhoA}
\end{align}
where $\mathbf A(\mathbf r)$ is the vector potential in the Coulomb gauge ($\nabla\cdot\mathbf A = 0$) for a static current source,
\begin{equation}
\mathbf A(\mathbf r) = \frac{\mu_0}{4\pi} \int \frac{\mathbf J(\mathbf r')}{\vert\mathbf r - \mathbf r'\vert}\ d\mathbf r'.\label{eq:AfromJ}
\end{equation}
\end{subequations}
It is easy to see that both expressions Eqs.~(\ref{eq:PemVJ}) and (\ref{eq:PemrhoA}) for $\mathbf P_{\mathrm{em}}$ are equivalent, by substituting Eq.~(\ref{eq:V}) into (\ref{eq:PemVJ}) and Eq.~(\ref{eq:AfromJ}) into (\ref{eq:PemrhoA}), and using $c^{-2} = \epsilon_0 \mu_0$.

First, we reconfirm that Eq.~(\ref{eq:P=Exm}) is valid for a magnetic dipole in the presence of a static electric field.\cite{furry}   The vector potential for a magnetic dipole at $\mathbf r_m$ in the Coulomb gauge is $\mathbf A = \mu_0 \mathbf m \times (\mathbf r-\mathbf r_m)/(4\pi \vert \mathbf r-\mathbf r_m\vert^3)$, so Eq.~(\ref{eq:PemrhoA}) gives
\begin{align}
\mathbf P_{\mathrm{em}}& = \mu_0\int d\mathbf r\ \rho(\mathbf r)\ \frac{\mathbf m\times (\mathbf r-\mathbf r_m)}{4\pi \vert \mathbf r-\mathbf r_m\vert^3} = \mu_0 \epsilon_0 \left[-\int d\mathbf r \frac{\rho(\mathbf r)\ (\mathbf r_m - \mathbf r)}{4\pi\epsilon_0 \vert \mathbf r-\mathbf r_m\vert^3}\right]\times \mathbf m,
\end{align}
which yields Eq.~(\ref{eq:P=Exm}) since the term in the square parentheses is the electric field at the position of the magnetic dipole and $\mu_0 \epsilon_0 = c^{-2}$.

\subsection{Expressions for $\mathbf P_{\mathrm{em}}$ for stationary electric dipole in a static magnetic field}

We now present four expressions for the total electromagnetic field momentum for an electric dipole $\mathbf p$ in the presence of a static current density $\mathbf J(\mathbf r)$ which produces a static magnetic field $\mathbf B(\mathbf r)$ and a corresponding vector potential in the Coulomb gauge $\mathbf A(\mathbf r)$.  In these expressions, the gradients and the magnetic fields are evaluated at the position $\mathbf r$ of the electric dipole.  The expressions are
\begin{subequations}
\begin{align} \mathbf P_{\mathrm{em}} &= (\mathbf p\cdot\nabla)\mathbf A(\mathbf r)\label{eq:Pem1}\\
 &= -\frac{\mu_0}{4\pi} \int \frac{[\mathbf p\cdot (\mathbf r -\mathbf r')]\;\mathbf J(\mathbf r')}{\vert\mathbf r-\mathbf r'\vert^3}\ d\mathbf r'\label{eq:Pem2}\\
 &= \mathbf B \times \mathbf p + \nabla(\mathbf p\cdot\mathbf A)\label{eq:Pem3}\\
 &=  \mathbf B \times \mathbf p - \frac{\mu_0}{4\pi}\int \frac{(\mathbf r-\mathbf r')\;[\mathbf p\cdot\mathbf J(\mathbf r')]}{\vert\mathbf r - \mathbf r'\vert^3}\ d\mathbf r'.\label{eq:Pem4}
\end{align}
\end{subequations}
Expression (\ref{eq:Pem1}), which was given in Ref.~\onlinecite{dipolesatrest}, can be derived  by taking a point dipole $\mathbf p$ at a position $\mathbf r$ to be the limit of point charges $-q$ at $\mathbf r$ and $q$ at $\mathbf r+\mathbf l$ in which $\vert\mathbf l\vert \equiv l \rightarrow 0$ and $q\rightarrow \infty$, with the product $\mathbf l q = \mathbf p$ being finite.  Using this in Eq.~(\ref{eq:PemrhoA}) and the expansion $\mathbf A(\mathbf r+\mathbf l) \approx \mathbf A(\mathbf r) + (\mathbf l\cdot \nabla)\mathbf A(\mathbf r)$, results in $\mathbf P_{\mathrm{em}} = \lim_{{\l\rightarrow 0}\atop{ql\rightarrow p}} q [\mathbf A(\mathbf r+\mathbf l)-\mathbf A(\mathbf r)] = \lim_{{\l\rightarrow 0}\atop{ql\rightarrow p}} (q\mathbf l\cdot\nabla)\mathbf A$, which gives expression (\ref{eq:Pem1}).

Combining Eq.~(\ref{eq:Pem1}) with Eq.~(\ref{eq:AfromJ}) and using the relationship (where $\nabla$ is the gradient with respect the variable $\mathbf r$)
\begin{equation}
\nabla \frac1{\vert\mathbf r - \mathbf r'\vert} = -\frac{\mathbf r - \mathbf r'}{\vert\mathbf r - \mathbf r'\vert^3},\label{eq:grad1/r}
\end{equation}
[or alternatively using the scalar potential for a point dipole, $V(\mathbf r) = \mathbf p\cdot\mathbf r/(4\pi\epsilon_0 r^3)$, in Eq.~(\ref{eq:PemVJ})] yields expression (\ref{eq:Pem2}).

Expression (\ref{eq:Pem3}) is obtained by using the vector identity [see, {\em e.g.}, Refs.~\onlinecite{griffiths} or \onlinecite{jackson}] $\nabla(\mathbf p\cdot \mathbf A) = \mathbf p \times (\nabla \times \mathbf A) + \mathbf A \times (\nabla \times \mathbf p) + (\mathbf p \cdot \nabla) \mathbf A + (\mathbf A\cdot\nabla) \mathbf p$, together with $\nabla\times \mathbf A  = \mathbf B$ and $\mathbf p$ not having any spatial dependence, gives expression (\ref{eq:Pem3}).
Finally, using Eq.~(\ref{eq:AfromJ}) in Eq.~(\ref{eq:Pem3}) and utilizing Eq.~(\ref{eq:grad1/r}) yields expression (\ref{eq:Pem4}).  Expressions (\ref{eq:Pem3}) and (\ref{eq:Pem4}) show that, in addition to the $\mathbf B \times \mathbf p$ term that is analogous to the $\mathbf E \times \mathbf m/c^2$ for a magnetic dipole in an electric field, there is an additional term which is dependent on the details of the current distribution which does not vanish in general.  Thus, the electromagnetic momentum of an electric dipole in the magnetic field {\em cannot} be expressed solely as a function of the dipole moment and the local magnetic field.

The differences between the forms of the expressions for $\mathbf P_{\mathrm{em}}$ for a magnetic dipole in an electric field and for an electric dipole in a magnetic field are due to the differences between the sources of static electric and magnetic field.  Static magnetic fields are caused by electric currents and static electric fields are caused by electric charges.   Therefore, despite the fact that the magnetic and electric fields for magnetic and electric dipoles have the same form (away from the dipoles themselves) the expressions for $\mathbf P_{\mathrm{em}}$ are different.    For more details, see Ref.~\onlinecite{dipolesatrest}.

\subsection{$\mathbf P_{\mathrm{em}}$ for various cases of electric dipoles in a magnetic field}\label{sec:Pem_for_various_cases}

We use the expressions Eqs.~(\ref{eq:Pem1}) -- (\ref{eq:Pem4}) to evaluate $\mathbf P_{\mathrm{em}}$ for a static electric dipole in a static magnetic field.  We obtain the correct result for the magnetic and dipole configuration shown in Fig.~1.  We then consider cases of an electric dipole in a uniform magnetic field produced by a spinning uniformly charged spherical shell and a uniform circular solenoid, where we reproduce the result $\frac12 \mathbf B \times \mathbf p$.\cite{mcdonald,gsponer,babson}  We see then that in the case of a uniform field produced by electric currents in two parallel plates, $\mathbf P_{\mathrm{em}}$ is not $\frac12 \mathbf B \times \mathbf p$. This is followed by a discussion of the reasons for the differences in $\mathbf P_{\mathrm em}$ in these cases.

\subsubsection{Configuration shown in Fig.~1}

We now re-evaluate $\mathbf P_{\mathrm{em}}$ for the example shown in Fig.~1.  The vector potential for a magnetic dipole $\mathbf m = m\hat{\mathbf z}$ at the origin in the Coulomb gauge, in spherical and cartesian coordinates, is\cite{griffiths}
\begin{align}
\mathbf A_{\mathbf m}(\mathbf r) &= \frac{\mu_0 \mathbf m \times \mathbf r}{4\pi r^3}   = \frac{\mu_0 m \sin\theta}{4\pi r^2}\, \hat{\boldsymbol\phi} = \frac{\mu_0 m}{4\pi} \frac{-y\hat{\mathbf x} + x\hat{\mathbf y}}{(x^2 + y^2 + z^2)^{3/2}}.
\end{align}
Therefore, for an electric dipole $\mathbf p = p\hat{\mathbf x}$ at position $R\hat{\mathbf x}$ [or in spherical coordinates, $r=R$, $\theta = \pi/2$ and $\phi = 0$] in the presence of the magnetic dipole at the origin, Eq.~(\ref{eq:Pem1}) gives
\begin{equation}
\mathbf P_{\mathrm{em}} =    p\left[\frac{\partial}{\partial r}\left(\frac{\mu_0 m\sin\theta}{4\pi r^2}\right)\right]_{r=R\atop{\theta=\pi/2}}\!\!\!\! \hat{\boldsymbol\phi}= - \frac{\mu_0 mp}{2\pi R^3}\,\hat{\boldsymbol\phi},\label{eq:Pem_corrected}
\end{equation}
which agrees with the result for $\mathbf P_{\mathrm{em}}$ using Eq.~(\ref{eq:P=Exm}), since $\hat{\boldsymbol\phi} = \hat{\mathbf y}$ when $\phi = 0$.  In this particular case, $\mathbf P_{\mathrm{em}} = 2\mathbf B \times \mathbf p$.

Alternatively, we can also use Eq.~(\ref{eq:Pem3}). In this case, it is more convenient to us the Cartesian-coordinate expression for $\mathbf A_{\mathbf m}(\mathbf r)$, which gives
\begin{equation}
\left[\nabla(\mathbf p \cdot\mathbf A_{\mathbf m})\right]_{(R,0,0)} =\frac{\mu_0 m p}{4\pi} \left[\nabla\frac{-y}{(x^2 + y^2 + z^2)^{3/2}}\right]_{(R,0,0)} = -\frac{\mu_0 m p}{4\pi R^3}\hat{\mathbf y} = \mathbf B \times \mathbf p.
\end{equation}
Using this in Eq.~(\ref{eq:Pem3}) also gives $\mathbf P_{\mathrm{em}} = 2\mathbf B \times \mathbf p$.

\subsubsection{Constant magnetic field inside uniformly charged spinning spherical shell}\label{subsec:spinsphere}

McDonald\cite{mcdonald}, Gsponer\cite{gsponer} and Babson {\em et al.}\cite{babson} evaluated $\mathbf P_{\mathrm{em}}$ for a capacitor or electric dipole $\mathbf p$, in a uniformly charged spinning spherical shell (or equivalently a uniformly magnetized sphere) producing a uniform field $\mathbf B$ inside the sphere.  In all cases, they obtained the result
$\mathbf P_{\mathrm{em}} = \frac12\mathbf B\times \mathbf p.$
We will see that this result is reproduced by the expressions given in the previous section, and is in fact independent of the position of the dipole (or equivalently capacitor) within the sphere.

The vector potential in the Coulomb gauge is\cite{griffiths,griffiths_short}
$\mathbf A_{\mathrm{in}}(\mathbf r) =
\frac13 {\mu_0 R\sigma} ({\boldsymbol\omega} \times \mathbf r)$ 
and this gives a uniform magnetic field inside the sphere of
$\mathbf B =
\frac23 \mu_0 \sigma R \boldsymbol\omega$.
Thus, for this case the vector potential inside the sphere can be written as
\begin{equation}
\mathbf A_{\mathrm{in}}(\mathbf r) = \frac12 \mathbf B\times \mathbf r.
\end{equation}

Since $\displaystyle \mathbf p \cdot \mathbf A_{\mathrm{in}} = \frac12 \mathbf p \cdot(\mathbf B \times  \mathbf r) = -\frac12 \mathbf r\cdot(\mathbf B\times \mathbf p)$, Eq.~(\ref{eq:Pem3}) gives
\begin{equation}
\mathbf P_{\mathrm{em}} = \mathbf B \times \mathbf p + \frac12 \nabla\left(\mathbf p\cdot (\mathbf B \times \mathbf r)\right) = \frac12 \mathbf B \times \mathbf p,\label{eq:Pemsphere}
\end{equation}
independent of the position of the electric dipole inside the sphere.  By linear superposition, this result holds for any distribution of dipoles within the sphere. This explains why in Ref.~\onlinecite{mcdonald} obtained result Eq.~(\ref{eq:Pemsphere}) for a the electromagnetic momentum of a spherical-cap capacitor inside a uniformly magnetized sphere (which is equivalent to a rotating uniformly charged spherical surface), where $\mathbf p$ is the total dipole moment of the charged capacitor.  In fact, so long as the capacitor is wholly contained within the sphere, the capacitor can any shape and the result would still hold.

\subsubsection{Constant magnetic field created by a cylindrical solenoid}

We now consider the case of an electric dipole in a region of constant magnetic field created by an infinite cylindrical solenoid of radius $R$ that is centerd along the $z$-axis.
The vector potential in the Coulomb gauge in cylindrical coordinates is\cite{grifA4cyl}
\begin{equation}
\mathbf A(\mathbf r) =
\begin{cases}
\frac12 B s \hat{\boldsymbol\phi} = \frac12\mathbf B \times \mathbf r & \text{if $s\leqslant R$},\\
\displaystyle\frac{BR^2}{2s} \hat{\boldsymbol\phi}, &\text{if $s\geqslant R$}
\end{cases}
\end{equation}
where $s$ is the distance from the $z$-axis.  These give $\mathbf B = B\hat{\mathbf z}$ inside the solenoid and $\mathbf B = 0$ outside.  We now evaluate $\mathbf P_{\mathrm{em}}$ for a dipole inside and outside the solenoid.

\underline{\sl Dipole within the solenoid}: Since the vector potential inside the cylinder is the same form as $\mathbf A_{\mathrm{in}}(\mathbf r)$ for the rotating sphere, the calculation also yields\cite{babson,mcdonald} $\mathbf P_{\mathrm{em}} = \frac12 \mathbf B \times \mathbf p$.   As in the case of the dipole within the sphere, this result is independent of the position of the dipole inside the solenoid, and therefore even a macroscopic object, such as a capacitor, with a total dipole moment $\mathbf p$ within the solenoid is also $\mathbf P_{\mathrm{em}} = \frac12\mathbf B\times \mathbf p$.

At this stage, one might surmise that even though $\mathbf P_{\mathrm{em}} = \frac12 \mathbf B \times \mathbf p$ does not hold for non-uniform magnetic fields, perhaps it is always true for uniform magnetic fields.  The next and subsequent examples show that this is not the case.

\underline{\sl Dipole outside the solenoid}: For a dipole outside the solenoid, the electromagnetic momentum is
\begin{align}
\mathbf P_{\mathrm{em}} &= (\mathbf p\cdot\nabla) \mathbf A = \frac{BR^2}2 \left(p_s \frac{\partial}{\partial s} + \frac{p_\phi}s \frac{\partial}{\partial \phi} + p_z\frac{\partial}{\partial z}\right) \frac{\hat{\boldsymbol\phi}(\phi)}s \nonumber\\
&= -\frac{BR^2}{2s^2} \left(p_s\hat{\boldsymbol\phi} + p_\phi\hat{\mathbf s}\right),\label{eq:Pemoutsol}
\end{align}
where $\partial\hat{\boldsymbol\phi}/d\phi = -\hat{\mathbf s}$ has been used.\cite{finn}
Since the magnetic field outside the solenoid is zero, clearly $\mathbf P_{\mathrm{em}} \ne \frac12 \mathbf B \times \mathbf p$.  In Appendix \ref{appendix:A1}, this result is reproduced by direct integration of the electromagnetic momentum density.

\subsubsection{Two parallel plates with counter-propagating currents}\label{sec:parallel_plates_current}

\begin{figure}
\includegraphics{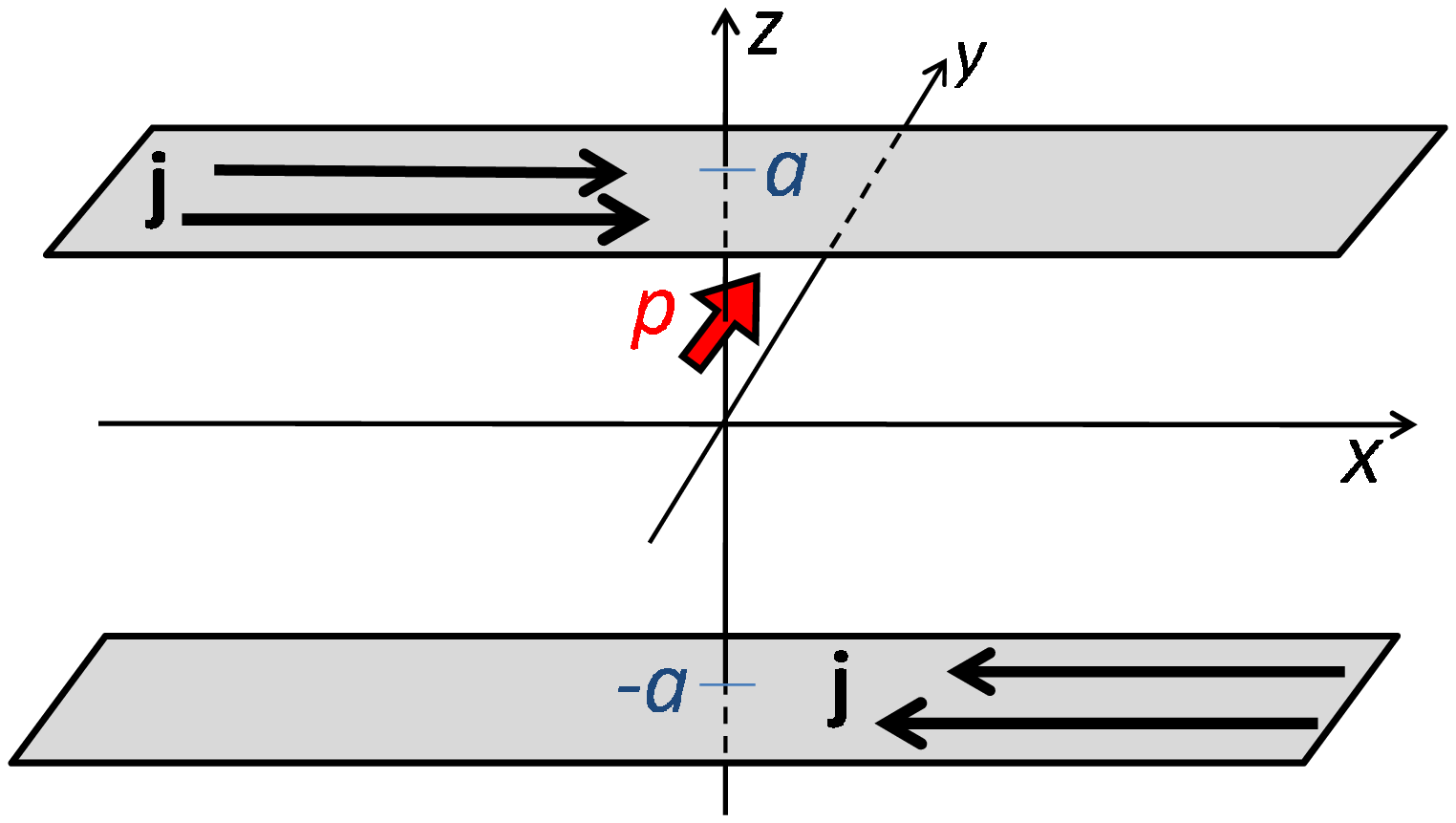}
\begin{caption}  {Electric dipole in a uniform magnetic field produced by currents in parallel plates.  The plates are in the $x$--$y$ plane at $z=\pm a$.  The current in the top (bottom) layer is in the positive (negative) $x$-direction, which produces a magnetic field in the positive $y$-direction.  The lateral extent of the plates is assumed to be much, much larger than the separation $2a$ of the plates.\label{fig:parallel_current}}\end{caption}
\end{figure}

Assume that there are a pair thin conducting plates are in the $x$--$y$ plane and they are at $z=\pm a$.   The plate on the top carries a uniform current density $j$ in the $+x$ direction and the plate at the bottom carries the an equal magnitude of current in the $-x$ direction, as shown in Fig.~\ref{fig:parallel_current}.  The calculation of the vector potential using Eq.~(\ref{eq:AfromJ}), is analogous to that of the scalar potential from parallel plate capacitors with a uniform charge distribution, and it gives a vector potential of
\begin{equation}
\mathbf A(\mathbf r) =
\begin{cases}
B z \hat{\mathbf x}, & \text{if $-a<z<a$},\\
\pm B a \hat{\mathbf x}, &  \text{if $z\gtrless a$},
\end{cases}
\end{equation}
where $B = \mu_0 j$, and a magnetic field of $B\hat{\mathbf y}$ for $\vert z\vert < a$ and $\mathbf B = 0$ for $\vert z\vert > a$.

The $\mathbf P_{\mathrm{em}}$ for an electric dipole outside the plates is zero, since $\mathbf A$ is constant.  For an electric dipole $\mathbf p$ in between the plates, using Eq.~(\ref{eq:Pem1}), the total electromagnetic momentum is
\begin{equation}
\mathbf P_{\mathrm{em}} = \left(p_x \frac{\partial}{\partial x} + p_y \frac{\partial}{\partial y} + p_z\frac{\partial}{\partial z}\right) B z \hat{\mathbf x} = p_z B \hat{\mathbf x}.\label{eq:Pemparallelplates}
\end{equation}
Thus, only the component of $\mathbf p$ perpendicular to the plates contributes to $\mathbf P_{\mathrm{em}}$.
This provides an example where there is a non-zero uniform magnetic field, but $\mathbf P_{\mathrm{em}}\neq\frac12\mathbf B\times \mathbf p$.

Note that in this case, even though the currents and hence the magnetic fields extend to infinity, the expression Eq.~(\ref{eq:PemrhoA}) is convergent and the surface term in Eq.~(\ref{eq:A1}) in Appendix \ref{appendix:A} goes to zero in the limit when the volume of integration goes to infinity.  Therefore, Eq.~(\ref{eq:Pem1}) [which is based on Eq.~(\ref{eq:PemrhoA})] can be used to obtain $\mathbf P_{\mathrm{em}}$.
In Appendix \ref{appendix:B}, the above result for $\mathbf P_{\mathrm{em}}$ is also obtained by evaluating $\epsilon_0\int d\mathbf r\ \mathbf E\times \mathbf B$ directly.

To circumvent any problems which might be associated with infinite currents and fields, one can assume that the current-carrying plates are large but finite, and the top and bottom plates are connected at the ends where the currents flow to and from.  In this case, there will be contributions to $\mathbf P_{\mathrm{em}}$ from the fringing fields, but these can be made arbitrarily small by increasing the size of the plates.   For readers who are skeptical that the above example is consequence of the unbounded currents and magnetic fields, which can sometimes give ill-defined physical quantities, an example of $\mathbf P_{\mathrm{em}} \ne \frac12\mathbf B \times \mathbf p$ for purely local currents in shown in the Appendix \ref{appendix:C}.

\subsection{Which gauge of $\mathbf A(\mathbf r)$ should be used?}

Eqs.~(\ref{eq:Pem1}) and (\ref{eq:Pem3}) indicate that $\mathbf P_{\mathrm{em}}$ for an electric dipole in a constant magnetic field is dependent on form of the vector potential $\mathbf A(\mathbf r)$ at the dipole.  In the case of a constant magnetic field $\mathbf B = B\hat{\mathbf y}$,  both $\mathbf A(\mathbf r) = \frac{B}2  \hat{\mathbf y}\times \mathbf r = \frac{B}2 (z\hat{\mathbf x}-x\hat{\mathbf z})$ and $\mathbf A(\mathbf r) =  B z\hat{\mathbf x}$ (or for that matter, $\mathbf A = B[ (1-\alpha) z \hat{\mathbf  x} - \alpha x\hat{\mathbf z}]$ for any constant $\alpha$) satisfy $\nabla \times \mathbf A= B\hat{\mathbf y}$ and $\nabla\cdot\mathbf A = 0$.   As shown above, the value of $\mathbf P_{\mathrm{em}}$ depends on which $\mathbf A(\mathbf r)$ is used.   So, which of these is the correct one to use?   The Helmholtz theorem states that the appropriate $\mathbf A$ in the Coulomb gauge ($\nabla\cdot\mathbf A =0$) is determined by $\mathbf B = \nabla\times \mathbf A$ over all space.  Provided that $\mathbf B$ goes to zero sufficiently fast as $r\rightarrow\infty$, $\mathbf A(\mathbf r)$ is determined by the current density $\mathbf J(\bf r)$ by Eq.~(\ref{eq:AfromJ}). This expression gives $\mathbf A = \frac{B}2  \hat{\mathbf y}\times \mathbf r$ for the interior regions of uniform cylindrical solenoids and spinning spheres, and $\mathbf A = \hat{\mathbf x} B z$ for interior region of parallel plates with constant current densities.\cite{Agaugenote}

\subsection{Contribution of the fringing electric fields of a capacitor to $\mathbf P_{\mathrm{em}}$}

A capacitor can be modeled as a superposition of point dipoles. Since the $\mathbf P_{\mathrm{em}}$ is linear in $\mathbf E$, which in turn is linear in $\mathbf p$, the results for $\mathbf P_{\mathrm{em}}$ for a point dipole are equally valid for a capacitor.   For cylindrical solenoids and spinning spheres of uniform surface charge, the factor of $\frac12$ in the result $\mathbf P_{\mathrm{em}} = \frac12\mathbf B \times \mathbf p$ implies $\mathbf P_{\mathrm{em}} = \frac12 \mathbf P_{\mathrm{capacitor}}$,  where $\mathbf P_{\mathrm{capacitor}}$ is the electromagnetic momentum that is stored in between the capacitor plates.  Obviously, the fringing electric fields of the capacitor contain a contribution of $-\frac12 \mathbf P_{\mathrm{capacitor}}$.
On the other hand, for a constant field generated by a ``solenoid" consisting of parallel conducing plates with counter-propagating currents, if the capacitor plates are oriented in the same way as the current-carrying plates, the $\mathbf P_{\mathrm{em}} = \mathbf P_{\mathrm{capacitor}}$, but if the capacitor plates are perpendicular to the plates carrying the current, $\mathbf P_{\mathrm{em}} = 0$.   What causes these differences in the total electromagnetic momentum?

\begin{figure}
\includegraphics{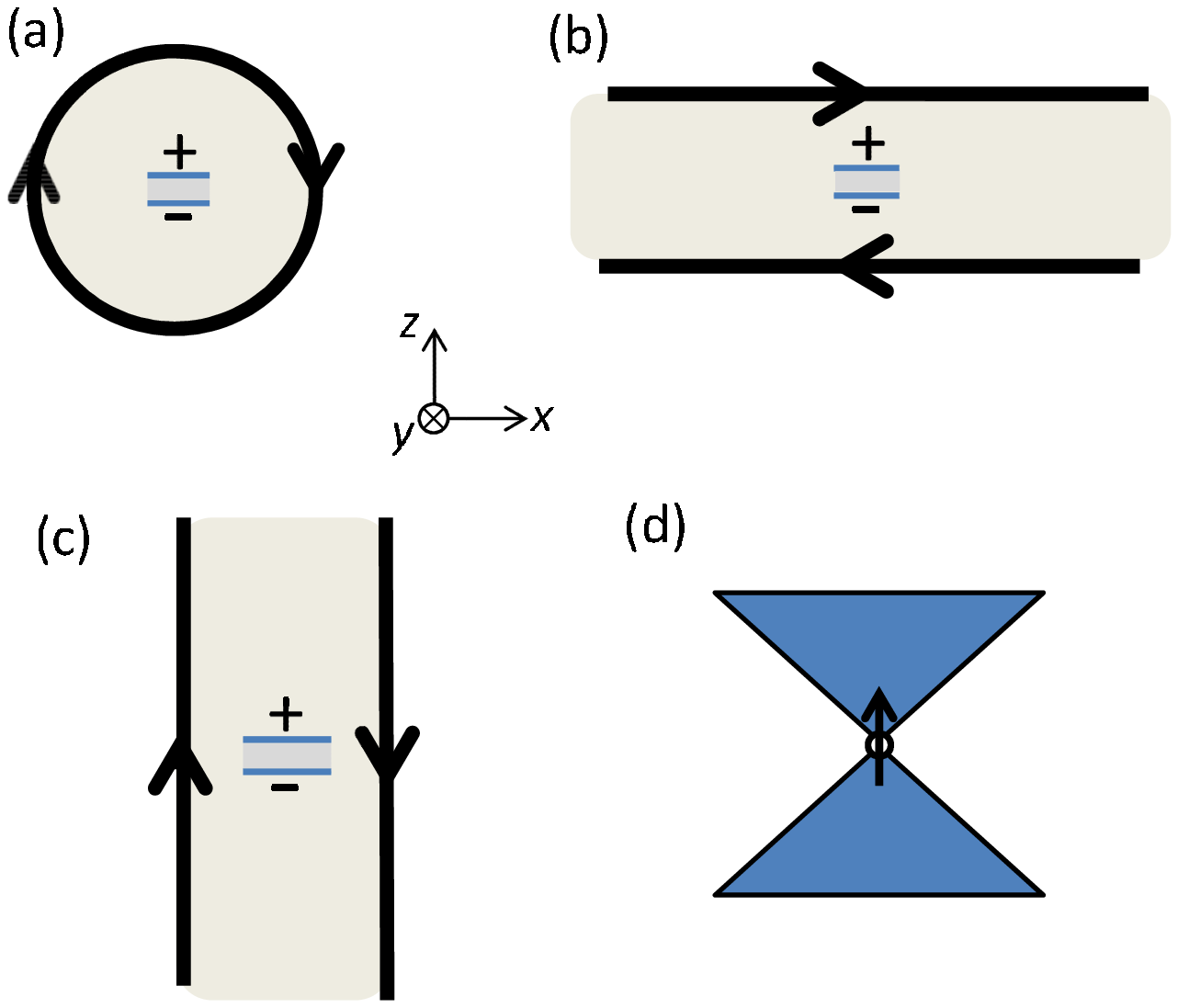}
\begin{caption} {Capacitors in the uniform magnetic fields produced by (a) infinite cylindrical solenoid with uniform current density around the circumference (b) infinite parallel plates with currents in opposite directions, and perpendicular the dipole of the capacitor and (c) same as (b) but with currents densities parallel and anti-parallel to the electric dipole moment of the capacitor.   In these three cases, the electromagnetic momenta $\mathbf P_{\mathrm{em}}$ is (a) $\frac12\mathbf B\times \mathbf p$, (b) $\mathbf B\times \mathbf p$ and (c) $0$, where $\mathbf p$ is the electric dipole moment of the capacitor and $\mathbf B$ is the magnetic field. Figure (d) shows the regions around an electric dipole that is oriented in the positive $z$-direction where the $z$-component of the electric field of the dipole is positive (shaded regions) and negative (white regions).  Thus, when the magnetic field is in the positive $y$-direction ({\em i.e.}, into the paper), the $x$-component of the electromagnetic momentum is negative in the shaded regions and positive in the white regions.  \label{fig:cap_Bfield}}
\end{caption}
\end{figure}

Consider the cases shown in Fig.~\ref{fig:cap_Bfield}(a) -- (c), in which the capacitors can be approximated as an electric dipole moment is the positive $z$-direction, and the magnetic field is in the positive $y$-direction.  In these cases, the $y$- and $z$-components of the total electromagnetic momentum $\mathbf P_{\mathrm{em}}$ are zero, so to determine the contribution to $\mathbf P_{\mathrm{em}}$, we need only consider the $z$-component of the electric field due to the dipole (since the electromagnetic momentum density is $\epsilon_0 \mathbf E \times \mathbf B$, and here $\mathbf B$ is in the $y$-direction.)  Fig.~\ref{fig:cap_Bfield}(d) indicates the regions where $E_z$ is positive and where it is negative around the electric dipole. Since the $z$-component of the electric field for a dipole at the origin is $E_{z}(\mathbf r) = \frac{p}{4\pi\epsilon_0 r^3} (3\cos^2\theta - 1) - \frac{p}{3\epsilon_0}\delta(\mathbf r)$, $E_z > 0$ in the cones that make an angle of $\cos^{-1}(1/\sqrt{3}) \approx 55^{\mathrm o}$ with respect to the positive and negative $z$-axes, as indicated by the shaded regions in Fig.~\ref{fig:cap_Bfield}(d).  When $\mathbf B$ in the positive $y$-direction, these regions will give a negative contribution to $P_{\mathrm{em},x}$, while conversely, the regions within approximately $35^{\mathrm{o}}$ of the ``equator" of the dipole, the electric field gives a positive contribution to $P_{\mathrm{em},x}$.

In the fringing field regions around the capacitors where the magnetic field exists, the volume of the regions where $E_z > 0$ ($E_z < 0$) increases (decreases) in the configurations shown in Figs.~\ref{fig:cap_Bfield}(b), (a) and (c), respectively.  This results in a corresponding increase in the contribution of the fringing fields, in that order.  In Fig.~\ref{fig:cap_Bfield}(b), apparently the fringing field contributions to $\mathbf P_{\mathrm{em}}$ of the  $E_z > 0$ and $E_z < 0$ regions cancel, and the result of the total electromagnetic momentum is the same as that contained within the capacitor plates.  Figs.~\ref{fig:cap_Bfield}(c) is at the other extreme of the cases considered here -- the regions where $E_z > 0$ in the fringing fields dominate to the extent that it completely cancels the electromagnetic momentum contained within the capacitor plates (where $E_z < 0$).   Figs.~\ref{fig:cap_Bfield}(a) is in between (b) and (c).


\section{Momentum imparted to the system when current or dipole is changed}\label{sec:III}

In this section, we attempt to clarify the discussion in Babson {\em et al.}\cite{babson} on the role of hidden momentum\cite{hiddenmomentum} in the conservation of total momentum of a system consisting of an electric dipole in a magnetic field, when the electromagnetic momentum is changed.  It is shown here that the mechanical impulse imparted to the system due to either a change in the current producing the magnetic field or the electric dipole is equal to the loss of electromagnetic momentum of the system, and hence it is not necessary to invoke hidden momentum to conserve the total momentum.

\subsection{Proof that impulse imparted to the system equals $-\Delta\mathbf P_{\mathrm{em}}$}

\subsubsection{Current is changed}

When the current is changed the vector potential changes by $\Delta \mathbf A(\mathbf r)$, which induces a change in the electric field $\mathbf E = -\partial\mathbf A/\partial t$ (the Faraday effect).  The momentum imparted to a charge distribution $\rho(\mathbf r)$ as the current changes is
\begin{align}
\vec{\mathcal I}_\rho &= \int dt \int d\mathbf r\ \rho(\mathbf r)\; \mathbf E(\mathbf r,t)\ 
= \int d\mathbf r\ \rho(\mathbf r)\left[- \int dt\ \frac{\partial\mathbf A}{\partial t}\right]\nonumber\\
&= -\int d\mathbf r\ \rho(\mathbf r)\ \Delta \mathbf  A(\mathbf r)
= -\Delta\mathbf P_{\mathrm{em}},
\end{align}
where the last equality comes from Eq.~(\ref{eq:PemrhoA}).Thus, when the magnitude of the current is changed, the change in the electromagnetic momentum is equal and opposite to the impulse given to the charges.\cite{EMmomargument}   Since this is true for any arbitrary distribution of charges, it is also true for the momentum imparted on the dipole.

\subsubsection{Electric dipole moment is changed}

When the dipole moment is changed, there is an electric current density $\mathbf J_{\mathrm{dip}}$ due to the transfer of charge within the dipole which experiences a Lorentz force $\mathbf F_{\mathrm{dip}} = \mathbf J_{\mathrm{dip}} \times \mathbf B$.  In addition, $\mathbf J_{\mathrm{dip}}$ itself produces a magnetic field, $\mathbf B_{\mathrm{dip}}$, which imparts a Lorentz force on the current in the solenoid.  Let us examine the impulses exerted by these forces and compare these to the change in the electromagnetic momentum.

Assuming the dipole is at the origin, the current distribution due to the change in the dipole moment is\cite{current_dipole}
\begin{equation}
\mathbf J_{\mathrm{dip}}(\mathbf r,t) = \dot{\mathbf p}(t) \delta(\mathbf r).
\end{equation}
The impulse on the dipole, which is equal to the momentum imparted on the dipole, is given by
\begin{align}
\vec{\mathcal I}_{\mathrm{dip}} &= \int dt\int \mathbf J_{\mathrm{dip}}(\mathbf r,t)\times \mathbf B(\mathbf r)\; d\mathbf r\nonumber\\
&= \int dt\ \dot{\mathbf p}(t)\times \mathbf B(\mathbf 0) = \Delta \mathbf p \times \mathbf B(\mathbf 0).\label{eq:impulse_on_dipole}
\end{align}


The current $\mathbf J_{\mathrm{dip}}(t)$ produces a magnetic field at position $\mathbf r'$ and time $t$ of\cite{dynamic_dipoles}
\begin{equation}
\mathbf B_{\mathrm{dip}}(\mathbf r',t) = -\frac{\mu_0}{4\pi} \left[\frac{\mathbf r' \times [\dot{\mathbf p} + (r'/c) \ddot{\mathbf p}]}{r'^3}  \right]_{\mathrm{ret}},
\end{equation}
where ``ret" indicates that $\dot{\mathbf p}$ and $\ddot{\mathbf p}$ are evaluated at retarded time $t_{\mathrm{ret}} = t - \vert\mathbf r'\vert/c$.   The impulse imparted on the current in the solenoid is
\begin{align}
\vec{\mathcal I}_{\mathrm{sol}} &= \int dt \int \mathbf J(\mathbf r') \times \mathbf B_{\mathrm{dip}}(\mathbf r',t)\ d\mathbf r'\nonumber\\
&=-\frac{\mu_0}{4\pi} \int \mathbf J(\mathbf r') \times \left(\frac{\mathbf r' \times \Delta\mathbf p}{r'^3}  \right)  d\mathbf r'\nonumber\\
&=\frac{\mu_0}{4\pi} \int \left[-\frac{\mathbf r'(\mathbf J(\mathbf r') \cdot\Delta\mathbf p)}{r'^3} + \Delta\mathbf p \left(\frac{\mathbf J(\mathbf r')\cdot\mathbf r'}{r'^3}\right) \right]d\mathbf r',\label{eq:16}
\end{align}
where the identity $\mathbf A \times (\mathbf B \times \mathbf C) = \mathbf B(\mathbf A\cdot\mathbf C) - \mathbf C(\mathbf A\cdot\mathbf B)$ has been used in the last equality.  (The $\ddot{\mathbf p}$ term in $\mathbf B_{\mathrm{dip}}(\mathbf r,t)$ does not contribute because $\Delta\dot{\mathbf p} = 0$, since $\mathbf p$ is constant before and after it is changed.)
The second term in the last expression in Eq.~(\ref{eq:16}) is zero because
\begin{align}
\int \frac{\mathbf J(\mathbf r')\cdot\mathbf r'}{r'^3}\ d\mathbf r' &= -\int \mathbf J(\mathbf r')\cdot\nabla'\left(\frac1{r'}\right)\ d\mathbf r'= \int \frac1{r'}\nabla'\cdot\mathbf J(\mathbf r')  - \nabla'\cdot\left(\frac{\mathbf J(\mathbf r')}{r'}\right) = 0
\end{align}
since $\nabla'\cdot\mathbf J(\mathbf r') = 0$ for static currents, and from the divergence theorem, $\int \nabla'\cdot\bigl(\mathbf J(\mathbf r')/r'\bigr) d\mathbf r' = \int_{\mathcal S} \mathbf J(\mathbf r')/r'\cdot d\mathbf a' = 0$, because $\mathcal S$ is a surface at infinity and $\mathbf J(\mathbf r')$ is be localized.
Therefore,
\begin{align}
\vec{\mathcal I}_{\mathrm{sol}} &= -\frac{\mu_0}{4\pi} \int \frac{\mathbf r' (\mathbf J(\mathbf r') \cdot\Delta\mathbf p)}{r'^3}\ d\mathbf r'.
\label{eq:18}
\end{align}
This implies that the total impulse given to the system due to the change $\Delta\mathbf p$ in the dipole moment at the origin is
\begin{equation}
\vec{\mathcal I}_{\mathrm{dip}} + \vec{\mathcal I}_{\mathrm{sol}} = \Delta \mathbf p \times \mathbf B(\mathbf 0) -\frac{\mu_0}{4\pi} \int \frac{\mathbf r' (\mathbf J(\mathbf r') \cdot\Delta\mathbf p)}{r'^3}\ d\mathbf r'.
\end{equation}
But from Eq.~(\ref{eq:Pem4}), this is exactly the negative of the change in the electromagnetic momentum for an electric dipole at $\mathbf r = 0$, so
\begin{equation}
\vec{\mathcal I}_{\mathrm{dip}} + \vec{\mathcal I}_{\mathrm{sol}} + \Delta\mathbf P_{\mathrm{em}} = 0,
\end{equation}
which shows that the momentum imparted to the system when the electric dipole moment changes is equal to the change in field electromagnetic momentum.

\subsection{Examples}

We now examine several cases of electric dipoles in magnetic fields created by different sources, specifically by a magnetic dipole, and by cylindrical, spherical and parallel-plate solenoids.  In each case, we show explicitly that when either the electric dipole or the magnetic-field-producing current is changed,  the impulse on the system is equal to the change in electromagnetic momentum.

\subsubsection{Magnetic and electric dipole at arbitrary orientation and displacement}

For an arbitrary placement of an magnetic moment $\mathbf m$ and an electric dipole $\mathbf p$ displaced by $\mathbf r\ne 0$ from the magnetic moment, the electromagnetic momentum is\cite{dipolesatrest}
\begin{equation}
\mathbf P_{\mathrm{em}} = \frac{\mu_0}{4\pi r^3} \left( (\mathbf m \times \mathbf p)-\frac3{r^2}(\mathbf p\cdot\mathbf r)(\mathbf m\times \mathbf r) \right)
 .\label{eq:Pem_for_dipoles}
\end{equation}
Let us assume that $\bm$ is at the origin and $\bp$ is at $\br$.

\underline{\em Magnetic dipole changed} -- When the magnetic dipole is changed by $\Delta\mathbf m$, the change in the electromagnetic momentum from Eq.~(\ref{eq:Pem_for_dipoles}) is
\begin{equation}
\Delta \mathbf P_{\mathrm{em}} = \frac{\mu_0}{4\pi r^3} \left( (\Delta\mathbf m \times \mathbf p)-\frac3{r^2}(\mathbf p\cdot\mathbf r)(\Delta\mathbf m\times \mathbf r) \right).\label{eq:DeltaP1}
\end{equation}
Let us assume, for notational simplicity, that the change in the dipole is quasi-static, so that one can ignore retardation effects and the $\ddot{\mathbf m}$ term.  (The derivation can easily be extended to include non-quasi-static changes.) This change produces an electric field\cite{dynamic_dipoles}
\begin{equation}
\mathbf E_{\mathrm{m.dip}}(\mathbf r,t) = \frac{\mu_0}{4\pi} \,\frac{\mathbf r \times \dot{\mathbf m}}{r^3}\;.\label{eq:E_dip_from_change_m}
\end{equation}
The force on the electric dipole is $\nabla(\mathbf p\cdot\mathbf E_{\mathrm{m.dip}})$, so the momentum imparted on the electric dipole due to a change in the magnetic moment is
\begin{equation}
\vec{\mathcal I} = \int \nabla(\mathbf p\cdot\mathbf E_{\mathrm{m.dip}})\ dt = \frac{\mu_0}{4\pi}\nabla\left(\frac{\mathbf p\cdot \mathbf r}{r^3}\right) \times \Delta\mathbf m = \frac{\mu_0}{4\pi} \left[\frac{\mathbf p}{r^3} - \frac{3(\mathbf p\cdot\mathbf  r)\mathbf r }{r^5}\right]\times \Delta\mathbf m.\label{eq:DeltaI1}
\end{equation}
Eqs.~(\ref{eq:DeltaP1}) and (\ref{eq:DeltaI1}) give $\Delta\mathbf P_{\mathrm{em}} + \vec{\mathcal I} = 0$.

It is easy to generalize the above from a point electric dipole to an arbitrary distribution of static charges $\rho(\mathbf r)$.  The electric field $\mathbf E_{\mathrm{dip}}(\mathbf r,t)$ induced by the change in $\mathbf m$ imparts an impulse on $\rho(\mathbf r)$ of (using Eq.~(\ref{eq:E_dip_from_change_m}), Coulomb's law for the electric field and
$\mu_0 \epsilon_0 = c^{-2}$)
\begin{align}
\vec{\mathcal I} &= \int d\mathbf r\ \rho(\mathbf r) \left[\int dt\ \mathbf E_{\mathrm{dip}}(\mathbf r,t)\right] =  \int d\mathbf r\ \frac{\mu_0}{4\pi} \left[\frac{\rho(\mathbf r)\,\mathbf r \times \Delta{\mathbf m}}{r^3}\right]\nonumber\\
& = -\frac1{c^2}\mathbf E \times \mathbf \Delta \mathbf m,
\end{align}
where $\mathbf E$ is electric field at the origin; {\em i.e.}, the position of the magnetic moment.  Since $\Delta \mathbf P_{\mathrm{em}} = c^{-2} \mathbf E \times \Delta\mathbf m$, this implies that $\vec{\mathcal I} + \Delta \mathbf P_{\mathrm{em}} = 0$ when $\mathbf m$ is changed for an arbitrary distribution of static charges.

\underline{\em Electric dipole changed} -- If the electric dipole is changed by $\Delta\mathbf p$,
the change in the electromagnetic momentum from Eq.~(\ref{eq:Pem_for_dipoles}) is
\begin{equation}
\Delta \mathbf P_{\mathrm{em}} = \frac{\mu_0}{4\pi r^3} \left( (\mathbf m \times \Delta\mathbf p)-\frac3{r^2}(\Delta\mathbf p\cdot\mathbf r)(\mathbf m\times \mathbf r) \right).\label{eq:DeltaP2}
\end{equation}
The impulse of the magnetic field on the charge current in the electric dipole, using Eq.~(\ref{eq:impulse_on_dipole}) and Eq.~(\ref{eq:dipoleB}) for the field of the magnetic dipole, is
\begin{equation}
\vec{\mathcal I}_{\mathrm{e.dip.}} = \Delta \mathbf p\times \mathbf B = \frac{\mu_0}{4\pi r^3}   \left[\frac{3(\mathbf m\cdot\mathbf r)(\Delta\mathbf p\times\mathbf r)}{r^2} - \Delta\mathbf p\times \mathbf m\right].\label{eq:B_dipole}
\end{equation}
In addition, the current of the electric dipole at $\mathbf r$, in the quasi-static limit, induces a magnetic field at $\mathbf r'$ of\cite{dynamic_dipoles}
\begin{equation}
\mathbf B_{\mathrm{e.dip}}(\mathbf r',t) = -\frac{\mu_0}{4\pi} \, \frac{(\mathbf r'-\mathbf r) \times \dot{\mathbf p}}{\vert\mathbf r'-\mathbf r\vert^3}\;.\label{eq:B_edip}
\end{equation}
The impulse on the magnetic dipole is
\begin{align}
\vec{\mathcal I}_{\mathrm{m.dip.}} &= \int \nabla'[\mathbf m\cdot\mathbf B_{\mathrm{e.dip}}(\mathbf r',t)]_{\mathbf r'=0}\ dt = -\frac{\mu_0}{4\pi} \nabla\left(\frac{\mathbf m\cdot[(\mathbf r'-\mathbf r)\times\Delta \mathbf p]}{\vert \mathbf r'-\mathbf r\vert^3} \right)_{\mathbf r'=0}\nonumber\\
&= -\frac{\mu_0}{4\pi} \nabla'\left(\frac{(\mathbf r'-\mathbf r)\cdot(\Delta \mathbf p\times\mathbf m)}{\vert \mathbf r'-\mathbf r\vert^3} \right)_{\mathbf r'=0}\nonumber\\
&= -\frac{\mu_0}{4\pi}\left[ \frac{ \Delta\mathbf p \times \mathbf m }{r^3}-  \frac{3\mathbf r\cdot(\Delta \mathbf p\times\mathbf m)\mathbf r}{r^5}\right].
\end{align}
Therefore, the total impulse on the system is
\begin{align}
\vec{\mathcal I} & = \vec{\mathcal I}_{\mathrm{e.dip.}} + \vec{\mathcal I}_{\mathrm{m.dip.}}\nonumber\\
&=  \frac{\mu_0}{4\pi r^5}   \left[3(\mathbf m\cdot\mathbf r)(\Delta\mathbf p\times\mathbf r) + 2(\bm \times \dbp) r^2 + 3\mathbf r\cdot(\Delta \mathbf p\times\mathbf m)\mathbf r\right]\nonumber\\
&=  \frac{\mu_0}{4\pi r^5}   \left[ - (\bm \times \dbp) r^2 + 3(\dbp\cdot\br) (\br\times\bm)\right]\label{eq:DeltaI2}
\end{align}
where in the last equality we use the identity\cite{identity}
\begin{equation}
(\bm\cdot\br)(\dbp\times\br) + (\br\cdot\br)(\bm\times\dbp) + (\dbp\cdot\br) (\br\times\bm) + [\br\cdot(\dbp\times\bm)]\br = 0. \label{eq:B1}
\end{equation}
Eqs.~(\ref{eq:DeltaP2}) and (\ref{eq:DeltaI2}) give
$\Delta \mathbf P_{\mathrm{em}} + \vec{\mathcal I} = 0.$

\subsubsection{Electric dipole on the axis of a cylindrical solenoid and at the center of a rotating charged spherical shell}

The cases in which an electric dipole is on the axis of a cylindrical solenoid and at the center of a rotating charged uniform spherical shell were considered by Babson {\em et al.}\cite{babson}.  In both cases, $\bP_{\mathrm{em}} = \frac12 \mathbf B \times \mathbf p$.  Furthermore the impulse imparted on the system when the dipole strength or the magnetic field is changed is also the same in both cases.

\underline{\em Magnetic field changed} --  When the current that is the source of the magnetic field is changed by $\Delta\mathbf B$, a straightforward generalization of the calculation given in Babson {\em et al.}\cite{babson} shows that the impulse given to the dipole is $-\frac12 \Delta\mathbf B \times \mathbf p$.  Since $\Delta\bP_{\mathrm{em}} = \frac12 \Delta \mathbf B \times \mathbf p$, this implies $\Delta\bP_{\mathrm{em}} + \vec{\mathcal I} = 0$.

\underline{\em Electric dipole changed} -- When the electric dipole is changed by $\dbp$, a straightforward generalization of the calculation of Ref.~\onlinecite{babson} shows that the impulse to the dipole is $\vec{\mathcal I}_{\mathrm{dip}} = - \mathbf B \times \Delta \mathbf p$ and the impulse to the solenoid is $\vec{\mathcal I}_{\mathrm{sol}} = \frac12 \mathbf B \times \Delta\bp $, giving a total impulse of $\vec{\mathcal I} = \vec{\mathcal I}_{\mathrm{dip}} + \vec{\mathcal I}_{\mathrm{sol}} = -\frac12 \mathbf B \times \Delta\bp $.  Again, $\Delta\bP_{\mathrm{em}}+\vec{\mathcal I}=0$.

\subsubsection{Electric dipole in a ``solenoid" of parallel counter-propagating currents}

In the case where the magnetic field is created by currents in parallel conducting plates in the configuration shown in Fig.~\ref{fig:parallel_current} and considered in Sec.~\ref{sec:parallel_plates_current}, the electromagnetic momentum is given by Eq.~(\ref{eq:Pemparallelplates}).

\underline{\em Magnitude of current in the solenoid is changed} -- When the current density is changed by $\Delta\mathbf j = \pm\Delta j\hat{\mathbf x}$ in the plate at $z = \pm a$, the magnetic field is changed by $\Delta \bB = \Delta B\;\hat{\mathbf y}
$ in between the plates, and therefore Eq.(\ref{eq:Pemparallelplates}) yields
\begin{equation}
\Delta\mathbf P_{\mathrm{em}} = p_z\;\Delta B\;\hat{\mathbf x}.
\end{equation}
The change in the magnetic field induces an electric field, which can be deduced by Amperian loops.  From the symmetry of the problem, the electric field must be in the $\hat{\mathbf x}$ direction, and independent of the $x$-coordinate.   Taking rectangular amperian loops which have a length $L$ in the $x$-direction and $d$ in the $z$-direction, the area through the loop is $L\,d$ and the change in flux $\Delta B L d$ is equal to $\int \mathbf E\cdot d\mathbf l$ around the loop.  This shows that $\int [E_x(z) - E_x(z+d)]\ dt = \Delta B d$.  
Thus, the momentum of a dipole inside the solenoid $\vert z \vert < a$, has momentum imparted of the form
\begin{equation}
\vec{\mathcal I} = \int q \left[E_x(z+d) - E_x(z)\right] dt\ \hat{\mathbf x} = -q d\; \Delta B\;  \hat{\mathbf x} = -\Delta B\; p_z \hat{\mathbf x}.
\end{equation}
This gives $\Delta \mathbf P_{\mathrm{em}} + \vec{\mathcal I} = 0$.

\underline{\em Electric dipole changed} --  When the electric dipole changes, Eq.(\ref{eq:Pemparallelplates}) gives
\begin{equation}
\Delta\mathbf P_{\mathrm{em}} =  \Delta p_z\;B\;\hat{\mathbf x}.\label{eq:Delta_p_Pem}
\end{equation}
The change in the $\mathbf p$ results in a current density $\mathbf j = \delta(\mathbf r-\mathbf r_{\mathrm{dip}}) \dot{\mathbf p}$ (where $\mathbf r_{\mathrm{dip}}$ is the position of the dipole) and $\int \dot{\mathbf p}\ dt = \Delta \mathbf p$.  As in previous instances considered, when $\mathbf p$ changes, there are two impulses on the system.  First, there is the Lorentz force of the magnetic field $\mathbf B$ on the current in the dipole, which gives an impulse of
\begin{equation}
\vec{\mathcal I}_{\mathrm{dip}} = \int dt \int d\mathbf r\ \mathbf j \times \mathbf B = \int dt\ \dot{\mathbf p} \times \mathbf B = \Delta\mathbf p\times \mathbf B.
\end{equation}  Then, there is the Lorentz force of the magnetic field induced by the change in the dipole (see Eq.~(\ref{eq:B_dipole})) on the currents in the parallel-plate solenoid.  We treat each of the three different components of the dipole, $\bp = p_x \hat{\mathbf x}$, $p_y \hat{\mathbf y}$ and $p_z\hat{\mathbf z}$, individually.
\begin{itemize}
\item[$\Delta p_z$:] When the $z$-component of $\mathbf p$ is changes by $\Delta p_z$, the impulse on the the dipole due to the magnetic field is $\vec{\mathcal I}_{\mathrm{dip}} = \Delta p_z\, B\,(\hat{\mathbf z}\times \hat{\mathbf y}) = - \Delta p_z\,B\,\hat{\mathbf x} $.
    The magnetic fields induced by the current in the dipole are in the azimuthal direction and symmetric with respect to the azimuthal angle, so by symmetry the total impulse $\vec{\mathcal I}_{\mathrm{sol}}$ due to these magnetic fields on the currents at $z=\pm a$ integrates to zero. Hence, $\vec{\mathcal I} = \vec{\mathcal I}_{\mathrm{sol}} + \vec{\mathcal I}_{\mathrm{dip}} = - \Delta p_z\,B\,\hat{\mathbf x} $.
\item[$\Delta p_y$:] When the $y$-component of $\mathbf p$ is changed by $\Delta p_y$, the impulse due to the magnetic field on the current in the dipole $\vec{\mathcal I}_{\mathrm{dip}} = \Delta \mathbf p \times \mathbf B = 0$.  By symmetry, the magnetic fields produced by the change in the electric dipole has $x$ and $z$ components.  The $x$-component of the magnetic field does not impart an impulse on the currents at $z = \pm a$, and the impulses from the $z$-components cancel due to symmetry.  Therefore, $\vec{\mathcal I} = \vec{\mathcal I}_{\mathrm{sol}} + \vec{\mathcal I}_{\mathrm{dip}} = 0$.
\item[$\Delta p_x$:] For $\Delta\mathbf p = \Delta p_x\hat{\mathbf x}$, the impulse on the current in the dipole is $\vec{\mathcal I}_{\mathrm{dip.}} = \mathbf p\times \mathbf B = \Delta p_x\, B (\hat{\mathbf x} \times \hat{\mathbf y}) = \Delta p_x\, B \hat{\mathbf z}$.   If we let $\mathbf r'$ be the position with respect to the dipole, then the current in the dipole produces a magnetic field $B_{\mathrm{e.dip}}(\mathbf r',t)$ given by Eq.~(\ref{eq:B_edip}) [with $\mathbf r = 0$ in that equation].  The current at $z=+a$ has a surface current density of $\mathbf j = j\hat{\mathbf x}$, so the force per unit area is $j\hat{\mathbf x} \times \mathbf B_{\mathrm{e.dip}}(\mathbf r',t)$.  The impulse on the entire sheet is
    \begin{align}
    \vec{\mathcal I}_{\mathrm{sol},+a} &= \int dx'\int dy' \int dt\ \mathbf j\times \mathbf B_{\mathrm{e.dip}}(\mathbf r',t) \nonumber\\
     &= \frac{\mu_0}{4\pi} \int dx'\int dy'\ \frac{j\hat{\mathbf x}\times (\hat{\mathbf x}\Delta p\times {\mathbf r'})}{r'^3}\nonumber\\
     &= \frac{\mu_0 j\ \Delta p}{4\pi} \int dx'\int dy'\ \frac{ \hat{\mathbf x}(\hat{\mathbf x}\cdot{\mathbf r'})-\mathbf r'}{r'^3}\nonumber\\
     &= -\frac{\mu_0 j\ \Delta p}{4\pi} \int dx'\int dy'\ \frac{y'\hat{\mathbf y} + z'\hat{\mathbf z}}{r'^3},
    \end{align}
    where $z'>0$ is the $z$-coordinate of the current sheet at $z=+a$ relative to the electric dipole.  The $\hat{\mathbf y}$ component vanishes because the integrand is anti-symmetric with respect to $y'$.  Changing variables of the integral $\int dx' \int dy'$ to $2\pi\int ds'\ s'$ (taking advantage of azimuthal symmetry of the integrand) gives, using a change of variables $\zeta = s'/z'$,
    \begin{align}
    \vec{\mathcal I}_{\mathrm{sol},+a} &= -\frac{\mu_0 j\ \Delta p}{2} \int_0^\infty  ds'\ s' \frac{z'}{(z'^2 + s'^2)^{3/2}}\hat{\mathbf x}\nonumber\\
    &= -\frac{B\ \Delta p}{2} \int_0^\infty d\zeta \frac{\zeta}{(1+\zeta^2)^{3/2}}\ \hat{\mathbf x} = -\frac{B\ \Delta p_x}{2}\hat{\mathbf x},
    \end{align}
    where we have used $B = \mu_0 j$.
    For the current in the plane at the $z=-a$, the calculation is identical except that the the current is $\mathbf j$ is in the opposite direction, and $z'<0$ so the change of variables $\zeta = s'/\vert z'\vert =  -s'/(-z')$ introduces a negative sign.  These two sign changes negate each other, so $\vec{\mathcal I}_{\mathrm{sol},-a} = -\frac12 B\,\Delta p\,\hat{\mathbf x}$, and the total impulse on the solenoid is
    \begin{equation}
    \vec{\mathcal I}_{\mathrm{sol}} = \vec{\mathcal I}_{\mathrm{sol},+a} + \vec{\mathcal I}_{\mathrm{sol},-a} = -B\; \Delta p_x\;\hat{\mathbf x}.
    \end{equation}
    Therefore $\vec{\mathcal I}_{\mathrm{sol}} + \vec{\mathcal I}_{\mathrm{dip}}= 0$.
\end{itemize}

Comparing these results for $\vec{\mathcal I}$ caused by $\Delta p_x$, $\Delta p_y$ and $\Delta p_z$ with the change in the field momentum Eq.~(\ref{eq:Delta_p_Pem}), we see that $\Delta\mathbf P_{\mathrm{em}} + \vec{\mathcal I} = 0$ for $\Delta\mathbf p$ in any direction.

\section{Discussion}\label{sec:IV}

An electric dipole in a static magnetic field and a magnetic dipole in a static electric field both generally give rise to a non-zero electromagnetic field momentum $\mathbf P_{\mathrm{em}} = \epsilon_0 \int d\mathbf r\ \mathbf E \times \mathbf B$.  While the $\mathbf P_{\mathrm{em}}$ for a point magnetic dipole in a static electric field can be expressed in terms of magnetic moment and the local electric field at the position of the moment, namely Eq.~(\ref{eq:P=Exm}), the same cannot be said for a point electric dipole in a static magnetic field; Eq.~(\ref{eq:P=Bxp}) is not true in general, even for locally uniform magnetic fields.   To determine $\mathbf P_{\mathrm{em}}$ for a point electric dipole in a static magnetic field, one needs to know the full current distribution which produces the magnetic field, or the appropriate derivatives of the magnetic vector potential in the Coulomb gauge at the position of the electric dipole; see Eqs.~(\ref{eq:Pem1}) -- (\ref{eq:Pem4}).   The difference between electric dipoles in static magnetic fields and magnetic dipoles in static electric fields is due the difference in the sources of static electric and magnetic fields; static electric fields are produced by charges whereas static magnetic fields are produced by charge currents.

It has also been shown that when either the electric dipole moment or the current producing the magnetic field is changed the mechanical impulse on the system $\vec{\mathcal I}$ is equal to the loss of electromagnetic momentum for an electric dipole in the magnetic field.  The motivation for showing this comes from Babson {\em et al.},\cite{babson} who stated in the conclusion section that ``there is no reason why this impulse should equal the momentum originally stored in the fields" in these situations, thus unintentionally inferring\cite{confusion} that it is necessary to take into account the hidden momentum\cite{PhysRev.171.1370,hiddenmomentum} in the system in order for the total momentum to be conserved.  Hidden momentum is a form of mechanical momentum that associated with internally moving parts in the presence of a potential gradient, and is a relativistic effect.  As shown in Ref.~\onlinecite{babson}, for a stationary electric dipole in a static magnetic field, the hidden momentum must be included for the total linear momentum to vanish, as is required since the``center-of-energy" is stationary.  However, there is no need to invoke hidden momentum in order to comply with the law of conservation of momentum when electromagnetic momentum is changed.\cite{errorinbabson}

It is interesting to note that, for a stationary electric dipole in a static magnetic field, when either the electric dipole moment or the current producing the magnetic field is changed, the overt mechanical momentum ({\em i.e.}, the non-hidden component of the mechanical momentum) is conserved, even when there is non-zero electromagnetic momentum initially present in the system.   By Newton's second law, the change in the mechanical momentum is equal to the impulse on the system; {\em i.e.}, $\Delta\mathbf P_{\mathrm{overt}} + \Delta\mathbf P_{\mathrm{hidden}} = \vec{\mathcal I}$.   Since $\vec{\mathcal I} = - \Delta \mathbf P_{\mathrm{em}}$, this implies that $\Delta\mathbf P_{\mathrm{overt}} = -\Delta[\mathbf P_{\mathrm{em}} + \mathbf P_{\mathrm{hidden}}]$.   But since $\mathbf P_{\mathrm{em}} = -\mathbf P_{\mathrm{hidden}}$ in a static system,\cite{hiddenmomentum} this gives $\Delta \mathbf P_{\mathrm{overt}} = 0$, independent of what the $\Delta\mathbf P_{\mathrm{em}}$ is.  
 
For example, considering a stationary electric dipole in the presence of a static magnetic field produced by a solenoid, if the current in the solenoid remains constant and dipole completely discharges, the dipole acquires momentum $\mathbf B \times \mathbf p$, and the solenoid always ends up with a momentum of $-\mathbf B \times \mathbf p$, regardless of what the initial $\mathbf P_{\mathrm{em}}$ is.  To illustrate this, consider the cases shown in Figs.~\ref{fig:cap_Bfield}(a) -- (c).  In Fig. 4(a), the hidden momentum that is initially contained in the solenoid is $-\frac12 \mathbf B \times \mathbf p$.  When the dipole is discharged, the impulse received by the solenoid from the magnetic field generated by the discharging dipole is $-\frac12 \mathbf B \times \mathbf p$.  In addition, the hidden momentum of $-\frac12\mathbf B \times \mathbf p$ that is initially in the solenoid is transformed into overt momentum when the dipole discharges, giving the solenoid a total overt momentum of $-\mathbf B \times \mathbf p$. In Fig. 4(b), when the dipole is discharged, the total impulse received by the solenoid currents is zero, but the solenoid had a hidden momentum of $-\mathbf B \times \mathbf p$ that is transformed into an overt mechanical momentum.  In Fig. 4(c), the total impulse received by the solenoid currents is $-\mathbf B \times \mathbf p$, but the solenoid had no hidden momentum.   On the other hand, if the dipole remains constant and the current in solenoid is turned off, the the dipole acquires an overt momentum that is equal to the initial electromagnetic momentum, and the solenoid acquires an overt momentum that is equal to the initial hidden momentum in the solenoid, which is equal to the negative of the initial electromagnetic momentum.   Thus, in all cases, the total change in the overt momentum of the dipole and solenoid is zero.\cite{DeltaPovert}


\begin{acknowledgements}

The hospitality of Prof.~Antti-Pekka Jauho at Aalto University, Helsinki, Finland, and Prof.~Christopher Stanton at the University of Florida, Gainesville, where some of this work was performed, and useful correspondence with Prof. David J. Griffiths are gratefully acknowledged.  The author also gratefully acknowledges an anonymous referee for useful suggestions that have improved the presentation of this manuscript.
\end{acknowledgements}

\appendix

\section{Derivation of Eq.~(\ref{eq:PemVJ})}\label{appendix:A}

To see under what conditions Eq.~(\ref{eq:PemVJ}) holds, let us first rederive this result for a finite volume $\mathcal V$, with surface $\mathcal S$.   For static electric and magnetic fields,
\begin{align}
\mathbf P_{\mathrm{em}} &= \epsilon_0 \int_{\mathcal V} d\mathbf r\ \mathbf E \times \mathbf B
 = -\epsilon_0\int_{\mathcal V} d\mathbf r\ \nabla V \times \mathbf B \nonumber\\ &= \epsilon_0\int_{\mathcal V} d\mathbf r \left[V (\nabla \times \mathbf B) - \nabla \times (V\mathbf B )\right]
= \frac1{c^2}\int_{\mathcal V} d\mathbf r\ V\mathbf J - \epsilon_0 \int_{\mathcal S} d\mathbf a\ V\;(\hat{\mathbf n}\times \mathbf B).\label{eq:A1}
\end{align}
where we have used $\mathbf E = -\nabla V$, $\nabla\times(V\mathbf B) = \nabla V \times \mathbf B + V \nabla\times\mathbf B$, $\nabla \times\mathbf B = \mu_0 \mathbf J$, $\mu_0\epsilon_0 = c^{-2}$, and $\int_{\mathcal V} d\mathbf r\ \nabla\times \mathbf A = \int_{\mathcal S} \hat{\mathbf n}\times \mathbf A$, where $\hat{\mathbf n}$ is the unit vector perpendicular to the surface. Letting $\mathcal V \rightarrow \infty$, one obtains Eq.~(\ref{eq:PemVJ}), provided that the last term in Eq.~(\ref{eq:A1}), $\epsilon_0 \int_{\mathcal S} d\mathbf a\ V\;(\hat{\mathbf n}\times \mathbf B)\rightarrow 0$ when $\mathcal S\rightarrow \infty$.  This condition is satisfied for an electric dipole in a uniform ``solenoid" of consisting of infinite parallel plates, since as the radius $r$ of the surface the radius of $\mathcal S$ increase, the integrand $V (\hat{\mathbf n} \times \mathbf B) \sim r^{-2}$ whereas area of $\mathcal S$ where the integrand is non-zero increases as $r$.   

\section{Derivation of Eq.~(\ref{eq:Pemoutsol}) by direct integration of the electromagnetic momentum density} \label{appendix:A1}

\begin{figure}
\includegraphics[scale=1.3]{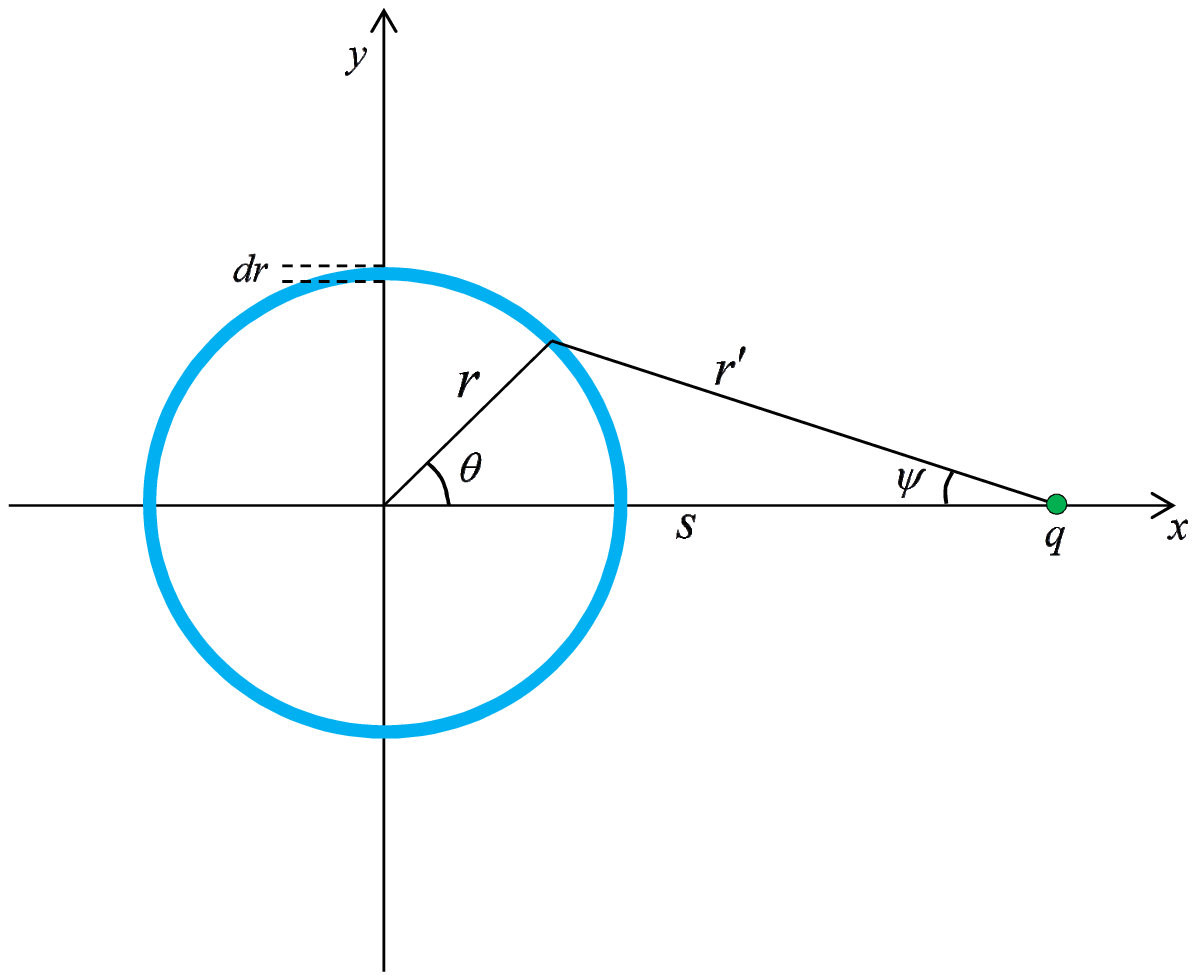}
\begin{caption} {Geometry of integration to obtain the electromagnetic momentum of a point charge outside a uniform solenoid.} \label{fig:integrate-cylinder}
\end{caption}
\end{figure}

We first derive the electromagnetic momentum for a point charge that is outside a uniform cylindrical solenoid.
The electromagnetic momentum of a point charge $q$ that is a perpendicular distance $s$ from the an infinitely long, infinitesimally small solenoid along the $z$-axis carrying magnetic flux $\Phi_m$ can be obtained by integrating the electromagnetic momentum over the length of the solenoid.  The result is, in cylindrical coordinates, is\cite{hu_EJP}
\begin{equation}
\mathbf P_{\mathrm{em}} = \frac{q \Phi_m}{2\pi s}\hat{\boldsymbol{\phi}}.\label{eq:Pemptchg}
\end{equation}
The $\mathbf P_{\mathrm{em}}$ for the charge $q$ a distance $s$ from the axis of of a solenoid of finite radius $R (<s)$ with uniform magnetic field $B$ can be obtained by using the above result and integrating over the cross section of the solenoid.   If we integrate along a shell of radius $r$ and width $dr$, as shown in Fig.~\ref{fig:integrate-cylinder}, by symmetry, the $\mathbf P_{\mathrm{em}}$ must be in the azimuthal direction.  The magnitude of the electromagnetic momentum due to this shell is
\begin{equation}
dP_{\mathrm{em}} = \frac{qB (r\, dr\,)}{2\pi}\,\int_{0}^{2\pi} \frac{\cos\psi}{r'}\ d\theta
\end{equation}
where $r'$, $\theta$ and $\psi$ are shown in Fig.~\ref{fig:integrate-cylinder}.  (The factor $\cos\psi$ arises because we are taking the azimuthal component of $d\mathbf \mathbf P_{\mathrm{em}}$.)  Using the law of cosines, which gives $r^2 = r'^2 + s^2 - 2 s r' \cos\psi$ and $r'^2 = r^2 + s^2 - 2 r s \cos\theta$, the integral can be evaluated to give
\begin{equation}
dP_{\mathrm{em}} = \frac{q B r\,dr}{4\pi s} \left(2\pi + \int_{0}^{2\pi} \frac{s^2 - r^2}{r^2 + s^2 - 2 r s \cos\theta} d\theta\right) = \frac{q B r\,dr}s.
\end{equation}
The total momentum is obtained by integrating from $r = 0$ to $r= R$, yielding
\begin{equation}
\mathbf P_{\mathrm{em}} = \frac{q B R^2}{2s}\hat{\boldsymbol{\phi}}.
\end{equation}

Consider now a dipole which consists of two charges $q$ and $-q$ separated by a distance $l$, in the limit $l\rightarrow 0$ and $ql\rightarrow p$.  For $\mathbf p$ in the $z$-direction, the contributions to $\mathbf{P}_{\mathrm{em}}$ from the two charges in the dipole cancel each other, so $\mathbf P_{\mathrm{em}} = 0$.  For $\mathbf p = p_s \hat{\mathbf s}$, given by charges $-q$ and $q$ at $s\hat{\mathbf s}$ and $(s+l)\hat{\mathbf s}$ respectively, Eq.~(\ref{eq:Pemptchg}) yields
\begin{equation}
\mathbf P_{\mathrm{em}} = \frac{ B R^2}{2}\hat{\boldsymbol\phi} \left[\lim_{{l\rightarrow 0}\atop{ql\rightarrow p_s}}
\left(-\frac{q}s+\frac{q}{s+l}\right) \right]= \frac{ B R^2}{2}\hat{\boldsymbol\phi} \lim_{{ql\rightarrow p_s}} \left[-\frac{ql}{s^2}\right] = -\frac{BR^2}{2s^2}p_s\hat{\boldsymbol\phi}.
\end{equation}
For $\mathbf p = p_\phi\hat{\boldsymbol\phi}$, given by charges $q$ and $-q$ are at the same distance $s$ from the $z$-axis but with azimuthal angles that are different by an angle $\Delta\phi = l/s$, Eq.~(\ref{eq:Pemptchg}) yields
\begin{align}
\mathbf P_{\mathrm{em}} &= \frac{B R^2}{2s} \lim_{{l\rightarrow 0}\atop{ql\rightarrow p_\phi}}\left[ q\hat{\boldsymbol\phi}(\phi+\Delta\phi)-q\hat{\boldsymbol\phi}(\phi)\right]
= \frac{B R^2}{2s}\lim_{ql\rightarrow p_\phi} \left[q\frac{l}s\frac{\partial\hat{\boldsymbol\phi}}{\partial\phi}\right]\nonumber\\
&= - \frac{BR^2}{2s^2}p_\phi \hat{\mathbf s}.
\end{align}
\section{Alternative derivation for $\mathbf P_{\mathrm{em}}$ due to electric dipole in between anti-parallel current sheets}\label{appendix:B}

The result Eq.~(\ref{eq:Pemparallelplates}) for the momentum of an electric dipole in a constant magnetic field created by conducting parallel plates can also be obtained by integrating $\epsilon_0 \mathbf E \times \mathbf B$ over all space.  To do this,  we divide the volume integral in between the parallel current-carrying plates into an integral across the entire $x$--$y$ plane, followed by an integral from $z = -a$ to $z = +a$; {\em i.e.},
\begin{subequations}
\begin{align}
\mathbf P_{\mathrm{em}} &= \epsilon_0\int_{-a}^a dz'\ \boldsymbol{\mathfrak E}(z') \times B\hat{\mathbf y};
\end{align}
where
\begin{align}
\boldsymbol{\mathfrak E}(z') &= \int_{-\infty}^\infty dx \int_{-\infty}^\infty dy\ \mathbf E(x,y,z').
\end{align}
\end{subequations}
is the integral of the electric field over the $x$--$y$ plane at $z=z'$.  Below, we evaluate $\boldsymbol{\mathfrak E}(z')$ for physical dipoles $\mathbf p = q\mathbf l$, with a finite $q$ and $l$, and then take the point dipole limit $q\rightarrow \infty$, $l\rightarrow 0$ and $ql\rightarrow p$ in the end.  We show here that $\boldsymbol{\mathfrak E}(z')$ is zero for any physical dipole except when plane at $z=z'$ lies in between the positive and negative charges of the dipole.

\begin{figure}
\includegraphics{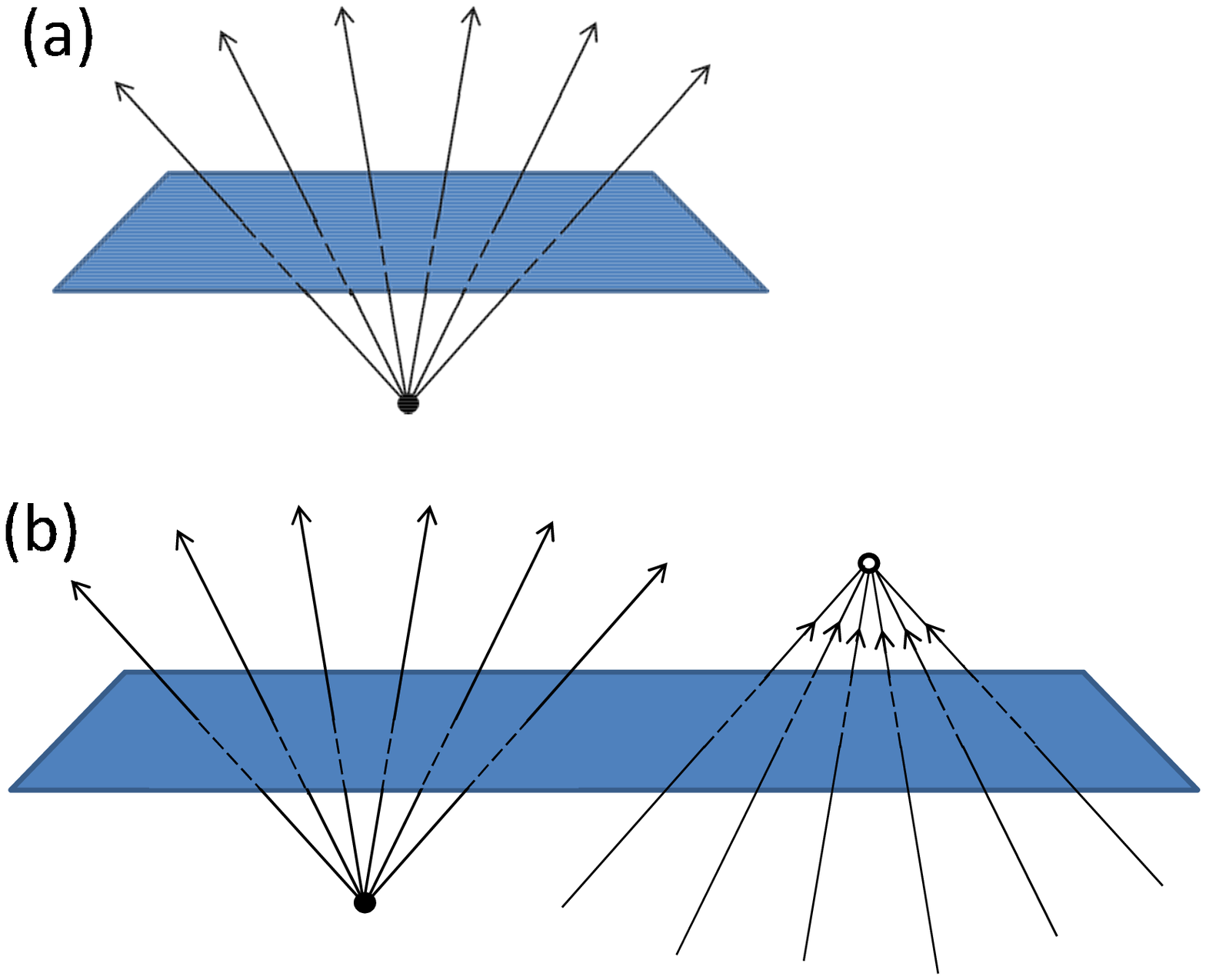}
\begin{caption}  {(a) Electric flux lines from a positive point charge piercing an infinite plane.  Since the plane is infinite, all the electric field lines with a component that is directed towards the plane will pierce the plane.  This implies that half the electric field lines that emanate from the point charge will pierce the plane.  Therefore, from Gauss' law, for an infinite plane (and a point charge $q$ which is not on the plane itself), the flux through the plane $\int \mathbf E\cdot d \mathbf r = \frac12 q/\epsilon_0$.  When there are two charges of equal magnitude and opposite sign on the {\em same} side of the plane, the the contributions of the $+$ and $-$ charges cancel each other and the electric flux through the plane is zero,.
(b) Electric flux lines for an infinite plane in between a positive (black dot) and negative (white dot) charges of equal magnitude.  The contribution for each charge to the flux is equal, and therefore the total flux is $q/\epsilon_0$.   \label{fig:EM_mom_Efield_pierce_plane}}\end{caption}
\end{figure}

By symmetry, the component $\boldsymbol{\mathfrak E}(z')$ along the surface of the plane due to a static point charge is be zero.  (For example, for a point charge on the $z$-axis, $E_{x,y}(x,y,z') = -E_{x,y}(-x,-y,z')$, so when $E_{x}$ or $E_{y}$ are integrated over the $x$--$y$, the contributions from $x,y$ and $-x,-y$ cancel each other.)  By linear superposition, this is also true for electric fields due to static dipoles.  Therefore, the only possible non-zero component is the component of $\mathbf E$ that is perpendicular to the surface, which is electric flux through the surface.    By Gauss' law,  the magnitude of the electric field flux through the plane due to a point charge $q$ is $\vert q\vert/(2\epsilon_0)$, independent of the distance of the charge from the plane.  (This is because half of the flux lines that emanate from the charge will pass through the plane -- see Fig.~\ref{fig:EM_mom_Efield_pierce_plane}(a).  When both the charges of a dipole are on the same side of the plane, the contributions from the positive and negative charges cancel.  The only case when $\boldsymbol{\mathfrak E}(z')$ is non-zero is when the plane straddles the charges; see Fig.~\ref{fig:EM_mom_Efield_pierce_plane}(b)).  So, if the dipole consists of a charge $-q$ at $(x_0,y_0,z_0)$ and $q$ at $(x_0+l_x,y_0 + l_y, z_0+l_z)$, then $\boldsymbol{\mathfrak E}(z')$ is equal to $-q/\epsilon_0\ \hat{\mathbf z}$ for $z_0 < z' < z_0 + l_z$, and zero otherwise.  Therefore,
\begin{equation}
\mathbf P_{\mathrm{em}} = \epsilon_0 \int_{z_0}^{z_0 + l_z} dz'\ \left(-\frac{qB}\epsilon_0\right) \ \hat{\mathbf z}\times \hat{\mathbf y}= q l_z B \hat{\mathbf x} = p_z B \hat{\mathbf x},
\end{equation}
reproducing Eq.~(\ref{eq:Pemparallelplates})

\section{Example of $\mathbf P_{\mathrm{em}} \ne \frac12\mathbf B \times \mathbf p$ for purely local currents}\label{appendix:C}

Current and charge configurations that extend to infinity can at times yield ill-defined results for quantities such as the electromagnetic field momentum.  To show that the results described in Sec.~\ref{sec:Pem_for_various_cases} are not artifacts of the unbounded currents used in the examples, in this appendix  a case is presented in which $\mathbf P_{\mathrm{em}} \ne \frac12\mathbf B \times \mathbf p$ even when only purely local currents and charges are present.

Consider a spinning sphere with a uniform surface charge, as in Refs.~\onlinecite{babson} and \onlinecite{mcdonald} and in Section \ref{subsec:spinsphere}, together with a torus which contains magnetic field. The magnetic field in the torus is caused by a surface current on the torus, and hence both current and field are local.  In the limit where the torus is infinitesimally small, the vector potential of the magnetic of the torus in the Coulomb gauge has the same form as the magnetic field of a point dipole;\cite{carron} namely, in spherical coordinates (with the $z$-axis perpendicular to the plane of the torus)
\begin{subequations}
\begin{align}
A_{\mathrm{tor},r} &= \frac{2C}{r^3} \cos\theta\\
A_{\mathrm{tor},\theta} &= \frac{C}{r^3} \sin\theta\\
C&= \frac{\mu_0 \mathcal V_{\mathrm{tor}}I}{16\pi^2}
\end{align}
\end{subequations}
where $\mathcal V_{\mathrm{tor}}$ is the volume of the torus and $I$ is the current times number of windings of the wire around the torus.

\begin{figure}
\includegraphics{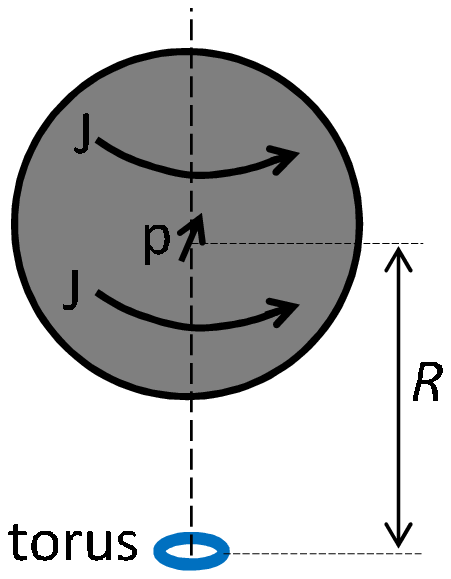}
\begin{caption}
{A situation where the electric current does not extend to infinity, and $\mathbf P_{\mathrm{em}} \ne \frac12 \mathbf B \times \mathbf p$, where $\mathbf B$ is the locally uniform magnetic field around electric dipole $\mathbf p$.  The electric dipole (indicated by the arrow) is in a uniform magnetic field created by a spinning sphere with a uniform surface charge, and is a distance $R$ away from and on the axis of a torus that contains magnetic flux.} \label{fig:torus}
\end{caption}
\end{figure}

The spinning sphere creates a uniform magnetic field $\mathbf B_{\mathrm{s}}$ inside the sphere, and the electromagnetic momentum due to the electric dipole and the sphere is $\mathbf P_{\mathrm{em,s}} = \frac12 \mathbf B_{\mathrm{s}}\times \mathbf p$.   For simplicity, let the dipole be at a distance $R$ from the torus and on the axis that passes through the torus, as shown in Fig.~\ref{fig:torus}.  The contribution to the electromagnetic momentum of the magnetic field inside the torus and the electric field of the dipole $\mathbf p = p_r\hat{\mathbf r} + p_\theta\hat{\boldsymbol\theta}$ (where the origin is taken to be the position of the torus), is $\mathbf P_{\mathrm{em,tor}} = C R^{-4} (-6 p_r\hat{\mathbf r} + p_\theta\hat{\boldsymbol\theta})$, by Eq.~(\ref{eq:Pem1}).  Therefore, the total electromagnetic momentum is $\mathbf P_{\mathrm{em,s}} + \mathbf P_{\mathrm{em,tor}} \ne \frac12 \mathbf B_{\mathrm{s}}\times \mathbf p$, in general.
\bibliographystyle{apsper}
\bibliography{EM_mom_bibl}

\begin{thebibliography}{27}
\expandafter\ifx\csname natexlab\endcsname\relax\def\natexlab#1{#1}\fi
\expandafter\ifx\csname bibnamefont\endcsname\relax
  \def\bibnamefont#1{#1}\fi
\expandafter\ifx\csname bibfnamefont\endcsname\relax
  \def\bibfnamefont#1{#1}\fi
\expandafter\ifx\csname citenamefont\endcsname\relax
  \def\citenamefont#1{#1}\fi
\expandafter\ifx\csname url\endcsname\relax
  \def\url#1{\texttt{#1}}\fi
\expandafter\ifx\csname urlprefix\endcsname\relax\def\urlprefix{URL }\fi
\providecommand{\bibinfo}[2]{#2}
\providecommand{\eprint}[2][]{\url{#2}}

\bibitem[{\citenamefont{Griffiths}(2010)}]{griffiths}
\bibinfo{author}{\bibfnamefont{D.~J.} \bibnamefont{Griffiths}},
  \emph{\bibinfo{title}{Introduction to Electrodynamics}}
  (\bibinfo{publisher}{Prentice-Hall, New Jersey}, \bibinfo{year}{2010}),
  \bibinfo{edition}{4th} ed.

\bibitem[{\citenamefont{Jacskon}(1999)}]{jackson}
\bibinfo{author}{\bibfnamefont{J.~D.} \bibnamefont{Jacskon}},
  \emph{\bibinfo{title}{Classical Electrodynamics}} (\bibinfo{publisher}{Wiley,
  New York}, \bibinfo{year}{1999}), \bibinfo{edition}{3rd} ed.

\bibitem[{mcd()}]{mcdonald}
\bibinfo{note}{K. T. Mcdonald, ``Electromagnetic momentum of a capacitor in a
  uniform magnetic field,"
  $<$www.hep.princeton.edu/$\sim$mcdonald/examples/cap\_momentum.pdf$>$ (June
  18, 2006; updated September 4, 2014; 14 pp).}

\bibitem[{gsp()}]{gsponer}
\bibinfo{note}{Andre Gsponer, ``On the electromagnetic momentum of static
  charge and steady current distributions," Eur. J. Phys. {\bf 28}, 1021--1042
  (2007).}

\bibitem[{bab()}]{babson}
\bibinfo{note}{David Babson, Stephen P. Reynolds, Robin Bjorkquist and David J.
  Griffiths, ``Hidden momentum, field momentum, and electromagnetic impulse,"
  Am. J. Phys. {\bf 77}, 862--833 (2009).}

\bibitem[{gri({\natexlab{a}})}]{griffithsresource}
\bibinfo{note}{David J. Griffiths, ``Resource Letter EM-1: Electromagnetic
  Momentum," Am. J. Phys. {\bf 80}, 7--18 (2012).}

\bibitem[{fur()}]{furry}
\bibinfo{note}{W. H. Furry, ``Examples of Momentum Distrubtion in the
  Electromagnetic Field and Matter," Am. J. Phys. {\bf 37}, 621--636 (1969).}

\bibitem[{fra()}]{franklin}
\bibinfo{note}{This was also pointed out in J. Franklin, ``The electromagnetic
  momentum of static charge-current distributions," Am. J. Phys. {\bf 82}, 869
  (2014). It should be noted that the present author does not agree with the
  main conclusions of Franklin's paper. See also D. J. Griffiths and V. Hnizdo,
  ``Comment on `The electromagnetic momentum of static-charge distributions,'
  by Jerrold Franklin [Am. J. Phys. 82, 869–875 (2014)]," Am. J. Phys. {\bf
  83}, 279 (2015), and J. Franklin, ``Response to `Comment on ``The
  electromagnetic momentum of static charge-current distributions" [Am. J.
  Phys. 83, 279 (2015)]' ", Am. J. Phys. {\bf 83}, 280 (2015).}

\bibitem[{gri({\natexlab{b}})}]{griffithsprivate}
\bibinfo{note}{David J. Griffiths, private communication.}

\bibitem[{hid()}]{hiddenmomentum}
\bibinfo{note}{For a review of hidden momentum, see
  Ref.~\onlinecite{griffithsresource} and references therein.}

\bibitem[{tho()}]{thomson}
\bibinfo{note}{J. J. Thomson, {\em Electricity and Matter}, (Charles Scribner's
  Sons, New York, 1904) pp. 30--33.}

\bibitem[{cal()}]{calkin}
\bibinfo{note}{M. G. Calkin, ``Linear Momentum of Quasistatic Electromagnetic
  Fields," Am. J. Phys. {\bf 34}, 921--925, (1966).}

\bibitem[{dip()}]{dipolesatrest}
\bibinfo{note}{David J. Griffiths, ``Dipoles at rest," Am. J. Phys. {\bf 60},
  979--987 (1992).}

\bibitem[{gri({\natexlab{c}})}]{griffiths_short}
\bibinfo{note}{David J. Griffiths, ``The field of a uniformly polarized
  object," Am. J. Phys., {\bf 60}, 187 (1992).}

\bibitem[{gri({\natexlab{d}})}]{grifA4cyl}
\bibinfo{note}{See {\em e.g.}, Ref.~\onlinecite{griffiths}, p.~247.}

\bibitem[{fin()}]{finn}
\bibinfo{note}{See, {\em e.g.}, John Michael Finn, {\em Classical Mechanics}
  (Infinity Science, Hingham, MA, 2008) p.~80.}

\bibitem[{Aga()}]{Agaugenote}
\bibinfo{note}{From the partial differential equation point of view, the
  equations $\nabla\cdot\mathbf A = 0$, $\nabla\times \mathbf A = \mathbf B$
  and $\nabla \times \mathbf B = \mu_0 \mathbf J$ are equivalent to $\nabla^2
  \mathbf A = -\mu_0 \mathbf J$; {\em i.e.}, each component of $\mathbf A$
  satisfies Poisson's equation. When the magnetic field is produced by surface
  currents the appropriate $\mathbf A$ is obtained by matching boundary
  conditions across each surface current in the same way that the scalar
  potential must be matched across surface charges. Each component of the
  magnetic vector potential must be continuous across the boundary, and for a
  surface current density $\mathbf K$, $\partial \mathbf
  A_{\|,\mathrm{in}}/\partial r_\perp - \partial\mathbf
  A_{\|,\mathrm{out}}/\partial r_\perp = \mu_0 \mathbf K$. See {\em e.g.},
  Section 5.4.2 in Ref.~\onlinecite{griffiths}.}

\bibitem[{EMm()}]{EMmomargument}
\bibinfo{note}{In fact, this argument was initially used to identify $\int
  \rho(\mathbf r) \mathbf A(\mathbf r)\ d\mathbf r$ (for $\mathbf A(\mathbf r)$
  in the Coulomb gauge) as the electromagnetic momentum. See
  Ref.~\onlinecite{calkin}.}

\bibitem[{cur()}]{current_dipole}
\bibinfo{note}{This can be derived in the following way. Assume that there is a
  charge $q$ at $l\hat{\mathbf z}$ and $-q$ at the origin, and the current
  flows from one charge to another along an infinitesimally small wire along
  the $z$-axis. The current density is $\mathbf J(\mathbf r) =
  \dot{q}\hat{\mathbf z} \delta(x)\delta(y) [\theta(l-z)\theta(z)] =
  (\dot{q}l)\hat{\mathbf z}\ \delta(x)\delta(y) [\theta(l-z)\theta(z)]/l$. Let
  $l\rightarrow 0$. Then, $[\theta(l-z)\theta(z)]/l \rightarrow\delta(z)$ and
  $\dot{q}l\hat{\mathbf z} = \dot{\mathbf p}$, so $\mathbf J = \dot{\mathbf
  p}\,\delta(\mathbf r)$.}

\bibitem[{dyn()}]{dynamic_dipoles}
\bibinfo{note}{David J. Grfiiths, ``Dynamic dipoles," Am. J. Phys. {\bf 79},
  867--872 (2011).}

\bibitem[{ide()}]{identity}
\bibinfo{note}{Proof: Applying $\br\cdot$ to the left hand side of
  Eq~(\ref{eq:B1}) and using $\br\cdot(\dbp\times\br) = (\br\times\bm)\cdot\br
  = 0$ gives \begin{equation} (\br\cdot\br)[(\bm\times\dbp)\cdot\br] +
  [\br\cdot(\dbp\times\bm)](\br\cdot\br) = 0 \end{equation} Applying
  $\br\times$ to the left hand side of Eq.~((\ref{eq:B1}) and using $\mathbf A
  \times (\mathbf B \times \mathbf C) = \mathbf B(\mathbf A\cdot\mathbf C) -
  \mathbf C(\mathbf A\cdot\mathbf B)$, gives \begin{align}
  &(\bm\cdot\br)[\br\times(\dbp\times\br)] +
  (\br\cdot\br)[\br\times(\bm\times\dbp)]+ (\dbp\cdot\br)
  [\br\times(\br\times\bm)] \nonumber\\ &=
  (\bm\cdot\br)[(\br\cdot\br)\dbp-(\dbp\cdot\br)\br] + (\br\cdot\br)
  [(\br\cdot\dbp)\bm - (\br\cdot\bm) \dbp]\nonumber\\ &+ (\dbp\cdot\br)
  [(\bm\cdot\br)\br-(\br\cdot\br)\bm] = 0.\nonumber \end{align} Since $\mathbf
  r\cdot\mathbf W = 0$ and $\mathbf r\times \mathbf W = 0 \Leftrightarrow
  \mathbf W = 0$ (for $\mathbf r\ne 0)$, this proves Eq.~(\ref{eq:B1}). (The
  identity can also be confirmed by the ``brute force" method of setting $\dbp
  = \Delta p \hat{\mathbf x}$, $\bm = m_x\hat{\mathbf x} + m_y\hat{\mathbf y}$
  and $\mathbf r = x\hat{\mathbf x} + y\hat{\mathbf y} + z\hat{\mathbf z}$,
  calculating the left hand side of Eq.~(\ref{eq:B1}) component by component
  and showing that each is zero.)}.

\bibitem[{con()}]{confusion}
\bibinfo{note}{The reader might draw the incorrect inference if the term ``this
  impulse," which refers to the impulse received by ``some element(s) in the
  system," is interpreted to mean the total impulse received by system.}

\bibitem[{Phy()}]{PhysRev.171.1370}
\bibinfo{note}{Sidney Coleman and J. H. Van Vleck, ``Origin of `Hidden Momentum
  Forces' on Magnets," Phys. Rev. {\bf 171}, 1370--1375 (1968).}

\bibitem[{err()}]{errorinbabson}
\bibinfo{note}{In fact, in Ref.~\onlinecite{babson}, Table I would have been
  correct if the the magnetic field was created by counter-propagating currents
  in parallel plates which are oriented in the same direction as the capacitor
  plates. (In this case, there is no net impulse on the currents when the
  dipole discharges.) Table II is inconsistent because it does not include the
  impulse of the magnetic field produced by the current of the discharging
  capacitor on the solenoid. When that is included, and the middle entry is
  relabelled ``Momentum delivered to capacitor and solenoid," all entries in
  the right hand column on that table would be $-\frac12 BQ\hat{\mathbf x}$.}

\bibitem[{Del()}]{DeltaPovert}
\bibinfo{note}{The fact that the total $\Delta \mathbf P_{\mathrm{overt}} = 0$
  can also be seen the in tables IV and V in Ref.~\onlinecite{babson}, which
  show the momentum transferred to the electric dipole and solenoid or spinning
  charged sphere when either the dipole moment vanishes or the current stops.
  The label $\mathbf p_{\mathrm{hid}}$ (for hidden momentum) in the second and
  third rows of these tables is somewhat misleading, because there is no longer
  any hidden momentum once the dipole discharges completely or the current
  stops. The hidden momentum would have been transformed to overt mechanical
  momentum. Thus, the entries in the middle two columns of the second and third
  rows of those tables give the overt mechanical momenta of the dipole and
  sphere/solenoid after the dipole discharges or the current stops. Note that
  in all cases the final total $\mathbf P_{\mathrm{overt}}$ is zero.}

\bibitem[{hu_()}]{hu_EJP}
\bibinfo{note}{Ben Yu-Kuang Hu, ``Introducing electromagnetic momentum," Eur.
  J. Phys. {\bf 33}, 873–-881 (2012).}

\bibitem[{car()}]{carron}
\bibinfo{note}{N. J. Carron, ``On the fields of a torus and the role of the
  vector potential," Am. J. Phys. {\bf 63}, 717--729 (1995).}

\end{thebibliography}

\newpage


\end{document}